\documentclass[11pt]{article}
\pdfoutput=1
\usepackage[utf8]{inputenc}
\usepackage{multirow}
\usepackage{amsmath, amsfonts, amssymb}
\usepackage{array}
    \newcolumntype{C}[1]{>{\centering\arraybackslash}m{#1}}
\usepackage{comment}
\usepackage{graphicx}
\usepackage{pdflscape}
\usepackage{psfrag}
\usepackage{amsthm}
\usepackage{enumerate}
\usepackage{arydshln}
\usepackage{pifont}
    
\usepackage{soul}
\usepackage{slashed}
\usepackage{subcaption}
\usepackage{mathrsfs}
\usepackage{ytableau}
 \usepackage{a4wide}
  \usepackage{tikz}
  \usepackage{tikz-cd}
  \usetikzlibrary{shapes.geometric}
  \usetikzlibrary{positioning}
  \usepackage{tcolorbox}
  \definecolor{dark-gray}{gray}{0.20}
  \definecolor{gray}{gray}{0.30}
  \definecolor{light-gray}{gray}{0.80}
  \definecolor{dark-red}{rgb}{0.7,0,0}
  \definecolor{dark-green}{rgb}{0.1,0.4,0}
  \definecolor{dark-blue}{rgb}{0.3,0.3,0.7}
  \definecolor{light-blue}{rgb}{0.8,0.8,1}
      \definecolor{swamp}{RGB}{240, 199, 197}
      
  \usepackage{pifont}

\usepackage{newunicodechar} 
\usepackage{setspace}

\usepackage{ifthen}
\renewcommand*{\arraystretch}{0.9}
\usepackage{longtable}

\newcommand{\be}{\begin{equation}}
\newcommand{\ee}{\end{equation}}
\newcommand{\eq}[1]{(\ref{#1})}

\def\be{\begin{equation}}
\def\ee{\end{equation}}
\def\bea{\begin{eqnarray}}
\def\eea{\end{eqnarray}}

\AtBeginDocument{

}

\numberwithin{equation}{section}

\usepackage{jheppub}

\usepackage{cleveref}

\hypersetup{
	colorlinks=true,
	linkcolor=dark-blue,
	citecolor=dark-red,
	urlcolor=dark-green,
	linktoc=page
}

\theoremstyle{definition}

\theoremstyle{remark}

\crefname{appendix}{Appendix}{Appendices}

\title{Stress-Testing Swampland Bounds\\ with Class $\mathcal{S}$ Theories}

\author{Gabriel Fenati$^1$,} 
\author{Miguel Montero$^{1,2}$}
\author{and Irene Valenzuela$^{1,2,3}$} 
\affiliation{$^1$ Instituto de F\'{i}sica Te\'{o}rica IFT-UAM/CSIC,
C/ Nicol\'{a}s Cabrera 13-15, Campus de Cantoblanco, 28049 Madrid, Spain}

\affiliation{$^2$ CERN, Theoretical Physics Department, 1211 Meyrin, Switzerland}
\affiliation{$^3$ Departamento de Física Téorica, Universidad Autónoma de Madrid, Cantoblanco, 28049 Madrid, Spain}
\emailAdd{gabriel.fenati@ift.csic.es, miguel.montero@csic.es, irene.valenzuela@cern.ch}
\abstract{We test the Sharpened Distance Conjecture (SharpDC) for CFTs, as well as the Refined Distance Conjecture (RDC), in AdS$_5$ gravitational spacetimes, using the CFT dual.
The former provides a lower bound on the exponential mass decay rate of the tower of states becoming light at infinite distance in the conformal manifold, while the latter bounds the bulk field range that can be traversed before the exponential behaviour of the tower kicks in.
We prove the SharpDC for any large $N$ Lagrangian supersymmetric gauge theory with a finite number $M$ of gauge factors and vanishing one-loop $\beta$-function. We also investigate whether it universally holds in the limits of large central charge, finding that it does \emph{not}: 
We exhibit concrete examples in Class $\mathcal{S}$ theories where the conjecture is violated in the $M\rightarrow\infty$ limit with $N$ fixed, where the size of the bulk internal geometry is much larger than the AdS curvature scale. 
We also test the RDC in the moduli space of Class $\mathcal{S}$ theories without punctures in the large $N$ limit, showing that it amounts to a precise upper bound on the Weil--Petersson diameter of the moduli space of Riemann surfaces of genus $g$. This bound is consistent with all the mathematical literature up to date, but so far remains unproven.}

\begin{document}
\emergencystretch 3em
\hypersetup{pageanchor=false}
\makeatletter
\let\old@fpheader\@fpheader
\preprint{IFT-26-119, 
CERN-TH-2026-201}

\makeatother

\maketitle

\hypersetup{
    pdftitle={},
    pdfauthor={},
    pdfsubject={}
}

\newcommand{\remove}[1]{\textcolor{red}{\sout{#1}}}

\newcommand{\red}[1][\text{(check)}]{\textcolor{red}{#1}}

\newcommand{\D}[1][\gamma]{\mathbf{D}_{\bf #1}}
\newcommand{\bvec}[1][\gamma]{\vec{b}_{\bf #1}}
\newcommand{\RFM}[2][T]{%
  \ifthenelse{\equal{#1}{T}}%
    {\frac{T^{#2}}{\Gamma}}%
    {\frac{\mathbb{R}^{#2}}{\mathcal{B}}}%
}
\newcommand{\Tr}[2]{{\rm Tr_{\bf #1}}(#2)}
\newcommand{\TrB}[1]{\Tr{B}{#1}}
\newcommand{\TrF}[1]{\Tr{F}{#1}}
\newcommand{\Vcas}{V_{\text{Cas}}}

\section{Introduction \& Conclusions}\label{sec:introductions}
The goal of the Swampland Program \cite{vafa2005stringlandscapeswampland} (see \cite{Brennan:2017rbf,Palti:2019pca,vanBeest:2021lhn,Grana:2021zvf,Harlow:2022ich,Agmon:2022thq} for reviews) is to chart and delimit the Landscape of consistent quantum theories of gravity. In doing so, we hope that some of the extrapolated features of the Landscape (encoded in the Swampland Conjectures) can be of use in phenomenological models of particle physics and cosmology. To achieve this, it is imperative to have as accurate a portrayal of the Landscape as possible; and to do so, it is essential to study and test Swampland Conjectures in all the solutions in the string Landscape, including Minkowski and Anti-de Sitter (AdS) vacua of String Theory.

Although, by now, there is quite a sizeable literature exploring Swampland Conjectures in AdS (see e.g. \cite{Lust:2019zwm} and \cite{ Perlmutter:2020buo}, for two prominent examples in the context of the Distance Conjecture \cite{Ooguri:2006in} in AdS), it is fair to say that many conjectures have not been subject to the same level of scrutiny as in flat space. Recently, a growing body of work has tested the Distance Conjecture in a variety of AdS/CFT settings \cite{Baume:2020dqd,Perlmutter:2020buo,Baume:2023msm,Ooguri:2024ofs, Calderon-Infante:2024oed, Calderon-Infante:2026rkj,Mantegazza:2026spd,Calderon-Infante:2026zmd,Baume:2026svx}. The evidence accumulated thus far suggests that the conjecture is a universal feature of unitary local CFTs, including those that do not admit an Einstein gravity dual. By contrast, considerably less is known about its refined versions, which seek to sharpen and quantify different aspects of the Distance Conjecture and are motivated primarily by results in flat space.
In this note we contribute to narrowing this gap, by studying two such refinements --- the Sharpened Distance Conjecture (SharpDC) \cite{Etheredge_2022} and the Refined Distance Conjecture (RDC) \cite{Baume_2016, Klaewer_2017, Rudelius:2023mjy} --- in interesting top-down families of $4d$ SCFTs. Although we explore general Lagrangian $4d$ CFTs, our more interesting results come from the study of Class $\mathcal{S}$ theories. This rich class of theories turns out to produce interesting challenges for both of these conjectures, showing that even supersymmetric corners of the Landscape still hold interesting treasures. 

We will first describe our results on the SharpDC, which bounds the rate at which the towers of states predicted by the Swampland Distance Conjecture become light in infinite distance limits. After reviewing the precise form of the statement for $4d$ SCFTs, we prove this conjecture for all large $N$ Lagrangian $4d$ SCFTs ($\mathcal{N}=1,2,4$) with $G=\prod_{a=1}^M G_a$ (each gauge factor is one among SU$(N)$, SO$(N)$, USp$(2N)$) with a finite number of gauge factors and at leading order in $N$. In addition, Einstein theories, which have $a\sim c$ at large $N$, saturate it. In the bulk interpretation, this implies that, for the non-Einstein theories under consideration, the tower originates from a string whose excitations become light always at a faster exponential rate than those of the perturbative critical ten-dimensional Type IIB string. Our proof goes beyond the bounds established thus far in the literature \cite{Calderon-Infante:2024oed,Calderon-Infante:2026rkj,Baume:2026svx} and can equivalently be recast as a proof that all the theories described above satisfy $c\geq a$ at leading order in $N$. We also study non-SUSY field theories with one-loop vanishing $\beta$-functions and show that these can violate the SharpDC. Constructing non-SUSY large $N$ CFTs is hard and them having a conformal manifold seems impossible, so perhaps this violation is not to be taken too seriously. 

As already noted in \cite{Perlmutter:2020buo,Calderon-Infante:2024oed}, we find that the SharpDC bound is satisfied only in the large $N$ limit. This suggests that, unlike the Distance Conjecture itself --- which also holds in finite-$N$ CFTs --- the SharpDC is a feature specifically associated with semiclassical gravity. We therefore investigate whether a parametrically large central charge, without a conventional large $N$ limit, is sufficient for the bound to hold. We find that it is not: the SharpDC is violated in theories in which the number of gauge-group factors becomes parametrically large but the gauge-group rank remains finite.
 More specifically, at finite $N$, we identify explicit families of exact SCFTs  with $c\rightarrow\infty$ which violate the Sharpened Distance Conjecture (equivalently, they have $a>c$; the perhaps non-trivial statement is that this behavior persists at large $M$). A simple example is the unique $4d$ $\mathcal{N}=2$ SCFT with SU$(2)^M$ gauge fields and trifundamental matter \cite{Gaiotto_2012}, in the limit $M\rightarrow\infty$, but there are others. 

The violations of the SharpDC we find are all obtained by sending gauge couplings to zero in a regime close to a  ``partial decompactification'', where part of the internal geometry is much larger than the AdS space. This regime is engineered in the CFT side by having a parametrically large number of gauge factors. This seems to indicate that the light string emerging in the overall weak-coupling limit has an exponential decay rate that is slower than what one would expect for the fundamental perturbative string (e.g. in $\mathcal{N}=4$ SYM). As mentioned above, violations of the SharpDC were already known to take place even in single-factor gauge theories for large but finite $N$, but there was no known example of a family of CFTs where the violation survived in the limit of large central charge $c\rightarrow\infty$; the examples in this note show that precisely this feature can arise close to partial decompactification limits.

On a related but somewhat orthogonal direction, we have also explored the exotic features that can be found for the moduli space of $A_{N-1}$ Class $\mathcal{S}$ theories without punctures, in the large $N$ limit where the theory admits an Einstein holographic dual. The moduli space metric receives significant corrections, and for large $N$, can be identified with the Weil--Petersson metric in the moduli space of Riemann surfaces of genus $g$. In this note, we provide a precise determination of the region in moduli space accurately captured by the Einstein region, and use it to test the Refined Distance Conjecture (RDC), which posits that one should not be able to traverse a large (geodesic) field range in Planck units before entering the asymptotic regime controlled by the tower of the Distance Conjecture, in the context of AdS compactifications. That the RDC holds is very important when it comes to phenomenological applications of the Distance Conjecture, e.g. in inflation (see e.g. \cite{ Baume_2016, Valenzuela_2017,Scalisi_2019}), since otherwise one could consider a scenario where the inflaton traverses a large distance in the bulk of moduli space, never being affected by the Swampland towers. 

For the theories that we study, we find that the RDC can be reformulated as an upper bound on the growth of the Weil--Petersson diameter of the moduli space of Riemann surfaces as a function of their genus: $\text{diam}(\mathcal{M}_g)\lesssim \sqrt{g}$. Interestingly, independent mathematical results show that $\text{diam}(\mathcal{M}_g)$ is \emph{lower} bounded by the same quantity! As a result, the RDC makes the mathematical prediction that the lower bound on $\sqrt{g}$ is saturated. While we have not seriously attempted to prove or falsify this bound, we emphasize that its failure would mean that we can find concrete top-down examples of bulk moduli spaces (in AdS) where an arbitrarily large field range can be traversed in Planck units without encountering any light states. One of the aims of this note is to catch the eye of an expert that might help us resolve this pressing question.

The note is structured as follows. In Section \ref{sec:SDCADS}, we review the version of the Sharpened Distance Conjecture bound in AdS and how the exponential mass decay rate is computed in the dual conformal field theory. We also clarify why we specify to the case of AdS$_5$/CFT$_4$.
In Section \ref{sec:checkSDC} we test the CFT SharpDC for both supersymmetric, in \ref{checksusy}, and non-supersymmetric, in \ref{checknonsusy}, large $N$ quiver Lagrangian CFTs. In \ref{checklargefactors} we provide counterexamples to the CFT SharpDC and we discuss its non-validity as a universal bound for large $c$ CFTs.
Finally, in Section \ref{sec:checkRDC}, after describing in \ref{reviewclassS} the nature of the different infinite distance limits of Class $\mathcal{S}$ theories, we address in \ref{testAdSRDC} the Refined Distance Conjecture in AdS and discuss its implication on the diameter of the moduli space of Riemann surfaces at large genus.

\section{The Sharpened Distance Conjecture in AdS}  \label{sec:SDCADS}
In this review section, we will recall the statement of the Sharpened Distance Conjecture \cite{Etheredge_2022} and how it applies to AdS compactifications. Consider a UV-completable effective field theory coupled to gravity, in generic $D$ spacetime dimensions. We will first consider the case where the vacuum is Minkowski, i.e. it has vanishing vacuum energy. Under these circumstances, the Swampland Distance Conjecture \cite{Ooguri:2006in} posits that an infinite tower of states becomes exponentially light asymptotically with the geodesic distance on the moduli space $\mathcal{M}_\phi$ of the theory,
\begin{equation}
    m\sim M_{D}\exp\{-\alpha\, \text{dist}_{\mathcal{M}_\phi}\}, \qquad\text{dist}_{\mathcal{M}_\phi}\longrightarrow\infty,
\end{equation}
where $M_D$ is the $D$-dimensional Planck mass.
This distance can be computed from the kinetic term of the massless scalar fields appearing in the EFT action, normalized in Planck units (for a review, see e.g. \cite{vanBeest:2021lhn}). The mass decay rate $\alpha$ is an $\mathcal{O}(1)$ number in Planck units which is central to the formulation of the conjecture. It controls how far can the excursion on moduli space  be before the tower becomes light and a breakdown of the effective field theory occurs. 

The Sharpened Distance Conjecture, originally introduced in \cite{Etheredge_2022}, posits a lower bound for $\alpha$. Specifically, it states that any infinite distance limit on the moduli space of a Minkowski QG theory features an infinite tower of states satisfying the Swampland Distance Conjecture, with
\begin{equation}
\alpha\geq\frac{1}{\sqrt{D-2}}\quad\quad\text{in $D$-dimensional Planck units.}
\end{equation}
We remark that this lower bound for $\alpha$ is valid for the lightest (i.e. leading) tower along any given infinite distance limit. In other words, this does not exclude the existence of other towers becoming light, in the same limits, with a mass decay rate $\alpha<\frac{1}{\sqrt{D-2}}$. 

In all known asymptotically flat compactifications, the decay rate of a tower of fundamental string modes saturates the SharpDC bound, $\alpha_\text{str}=1/\sqrt{D-2}$ when behaves as the leading tower. By the Emergent String Conjecture (ESC) \cite{Lee_2018}, all infinite distance limits in flat space quantum gravity correspond to either decompactification or to a critical string becoming tensionless in Planck units. In the first case, if we consider a decompactification from $D$ to $D+n$ dimensions, the leading KK tower becomes light with an exponential rate\footnote{In highly warped setups where the decompactification limit bring us to a running time-dependent solution, the asymptotic KK exponential rate can be modified and take a smaller value \cite{Etheredge:2023odp,Raucci:2026fzp}.}
\begin{equation}\alpha_\mathrm{KK}=\sqrt{\frac{n+D-2}{n(D-2)}}\end{equation}
along the direction of steepest descent of its mass.
Hence, putting together the SharpDC and the ESC, one has the following bounds for the mass decay rate of the leading tower along any infinite distance limit \cite{Etheredge_2022}
\begin{equation}
    \frac{1}{\sqrt{D-2}}\leq\alpha\leq\sqrt{\frac{D-1}{D-2}},
\end{equation}
where the upper bound is the maximal value for $\alpha_\mathrm{KK}$ (obtained when $n=1$). The validity of these bounds has been checked in many examples of flat compactifications \cite{Etheredge_2022, Aoufia:2025ppe, Etheredge:2025ahf, Kaufmann:2024gqo, Etheredge:2024amg, Etheredge:2024tok, Castellano:2023jjt, Castellano:2023stg, Rudelius:2023odg, Etheredge:2023usk, Etheredge:2023odp,Raucci:2026fzp}.
\medskip

The focus of this note is, however, in AdS quantum gravity. In this case, one still expects the Distance Conjecture to hold, and to have towers of states becoming light in infinite distance limits. There are two qualitative infinite distance limits one can take \cite{Calderon-Infante:2024oed}: large $N$, where the limiting procedure involves a discrete parameter, or limits in the moduli space of the AdS vacuum at fixed $N$, on which we will focus and which have also seen a lot of recent progress \cite{Baume:2020dqd,Perlmutter:2020buo,Baume:2023msm,Ooguri:2024ofs,Calderon-Infante:2026rkj,Mantegazza:2026spd,Calderon-Infante:2026zmd,Baume:2026svx}.
The AdS moduli space, via the AdS/CFT correspondence, maps to the conformal manifold $\mathcal{M}_\mathrm{CFT}$ of the dual conformal field theory. For $d>2$ dimensional CFTs a version of the Distance Conjecture directly on CFT language was formulated in \cite{Perlmutter:2020buo, Baume:2020dqd} and partially proved in \cite{Baume:2023msm}. It states that any local and unitary CFT located at a point at infinite Zamolodchikov distance \cite{Zamolodchikov:1986gt} on the conformal manifold has in its spectrum an infinite tower of higher-spin (HS) operators\footnote{That is a tower of operators with spin $J>2$ and increasing.} saturating the unitarity bound, such that their anomalous dimension goes to zero,
\begin{equation}
    \gamma_J\equiv\Delta_J-\Delta^{\text{(unit)}}_J \longrightarrow 0 , 
\end{equation}
exponentially with the distance measured by the Zamolodchikov metric. This means that a (sub)sector of the CFT becomes free in the limit, as higher-spin currents become conserved \cite{maldacena2011constrainingconformalfieldtheories}. If we consider large $N$ CFTs, then we can translate this statement into a statement for the dual weakly-coupled (i.e. semiclassical) gravitational theory on AdS.\footnote{By weakly-coupled we mean that $\ell_\mathrm{AdS}\,M_{d+1}\gg1$, where $\ell_\mathrm{AdS}$ is the AdS radius and $M_{d+1}$ is the $D=d+1$ dimensional Planck scale.} At large $N$, the infinite tower is formed by single-trace HS operators. 
A single-trace operator of scaling dimension $\Delta$ and spin $J$ of the CFT$_d$ maps to a state in AdS$_{d+1}$ with same spin and mass given by\begin{equation}
    (m\,\ell_\mathrm{AdS})^2=(\Delta+J-2)(\Delta-J-d+2),
\end{equation}
so the tower of higher-spin operators becomes a tower of bulk HS states with mass scale 
\begin{equation}    \label{towerads}
    m_J\sim M_{d+1}\exp\{-\alpha\, \text{dist}_{\mathcal{M}_\phi}\},\qquad \text{dist}_{\mathcal{M}_\phi}\longrightarrow\infty.
\end{equation}
 We see that the tower mass scale goes to zero exponentially just as in the flat space counterpart. The tower of HS states is naturally interpreted as the excitation states of a string becoming light in Planck units. An important difference with the flat space version of the conjecture is that in AdS with $d>2$ we only seem to have string (HS) towers at infinite distance on the moduli space, and there are no KK (decompactification) limits in more than two dimensions \cite{Perlmutter:2020buo, Calderon-Infante:2024oed}.
 
 Another qualitative difference from the flat space case is that the AdS curvature scale $\ell_\mathrm{AdS}$ provides a second lengthscale in the problem. Since the string is becoming light, when the characteristic lengthscale of the string $\ell_\mathrm{str}$ becomes comparable to $\ell_\mathrm{AdS}$, which determines the spacetime curvature, the spacetime geometry ceases to be accurately described by Einsteinian theory. Therefore, close enough to the infinite distance limits, the Einstein theory breaks down, unlike in the flat space case where the Einsteinian description remains valid at arbitrarily large distances, provided that one focues in the deep infrared.\footnote{For instance, little string theories in flat space produce local CFTs at low energies. The same construction in an AdS compactification leads to non-critical strings \cite{Calderon-Infante:2026zmd}.}
 
 Although the CFT Distance Conjecture, as a statement, makes sense in any dimension and with or without supersymmetry, in this note we will follow previous work \cite{Baume:2020dqd,Perlmutter:2020buo,Baume:2023msm,Calderon-Infante:2026rkj,Mantegazza:2026spd,Calderon-Infante:2026zmd,Baume:2026svx}
 and we will only consider  $4d$ SCFTs. This is because there is no known example of a non-SUSY conformal manifold in $d>2$, and the case $d=4$ is the only one where one can take infinite distance limits on $\mathcal{M}_\mathrm{CFT}$ (see \cite{Perlmutter:2020buo} and references therein). In all known examples, all infinite distance limits on the conformal manifold $\mathcal{M}_\mathrm{CFT}$ admit a description in terms of a weakly-coupled gauge theory (which is, of course, conformal). More concretely, as we approach the infinite distance limit, the theory can be described as non-abelian gauge theory with group $G$, possibly coupled to strongly-coupled matter, and the diverging modulus controls a gauge coupling of a subsector of the theory. In the infinite distance limit, a subgroup $G_\mathrm{free}\subseteq G$ of the gauge sector becomes free and the Zamolodchikov metric in these regions develops a cusp as \cite{Perlmutter:2020buo}
\begin{equation}
    ds^2_\mathrm{Zam}\simeq 24\dim G_\text{free}\frac{d\tau\,d\bar\tau}{(\text{Im }\tau)^2},\qquad \text{Im }\tau\longrightarrow\infty,
\end{equation}
where the complexified gauge coupling $\tau\equiv\frac{\theta_\mathrm{YM}}{2\pi}+\frac{4\pi i}{g_\mathrm{YM}^2}$ parametrizes the conformal manifold (here one-dimensional for simplicity).

Having an explicit gauge theory Lagrangian description of the infinite distance limit allows one to compute the value of $\alpha$ explicitly. The corresponding bulk mass decay rate is given by \cite{Perlmutter:2020buo}
\begin{equation}    \label{alpha}
    \alpha = \sqrt{\frac{2\,c}{\dim G_\mathrm{free}}},
\end{equation}
where the central charge $c$ \cite{Myers_2011} appears due to the change from AdS to bulk Planck units. It then depends on the dimension of the factor of the gauge sector becoming free in the given limit, and on the degrees of freedom of the whole CFT via the central charge $c$.

Once we have \eqref{alpha}, we can use it to compute its value in many examples, thus providing a landscape of values of $\alpha$.  Reference \cite{Perlmutter:2020buo,Calderon-Infante:2024oed} did precisely this, considering all Lagrangian $4d$ $\mathcal{N}\geq 1$ SCFTs, with a conformal manifold, a \textit{simple} gauge group $G$, and admitting a large $N$ expansion (hence $G=$ SU$(N)$, SO$(N)$, USp$(2N)$). 
The authors carried out a systematic classification in terms of the exponential rate $\alpha$ corresponding to weak-coupling limits. We remark that their computation is performed in the perturbative CFT regime, where one first fixes $N$ to a large value, and then sends the gauge coupling $g_\mathrm{YM}$ to zero. In terms of the 't Hooft coupling $\lambda$, this means 
\begin{equation}
    \lambda\equiv g^2_\mathrm{YM}\,N=g_s\,N\ll1.
\end{equation}
They found $\alpha\geq 1/\sqrt{2}$ in all the cases. More precisely, they found three universality classes, with values 
\begin{equation}\alpha = \frac{1}{\sqrt{2}}, \sqrt{\frac{7}{12}}, \sqrt{\frac{2}{3}}.\end{equation}
As studied in  \cite{Calderon-Infante:2024oed}, all theories in each class share the same ratio of the central charges $a$ and $c$ as well as the same Hagedorn density of high-energy states. Therefore, they were identified as three different types of tensionless strings in the bulk, an expectation that was confirmed in \cite{Calderon-Infante:2026zmd} by studying their microscopic nature via the construction of their dual brane configuration. It turned out that $\alpha=\frac{1}{\sqrt{2}}$ (the smallest value) is a tower of $10d$ Type IIB critical string modes, whereas $\alpha=\sqrt{\frac{2}{3}}$ is a tower of subcritical string modes (propagating in a background obtained from taking a double-scaling limit of two NS5-branes) and  $\alpha=\sqrt{\frac{7}{12}}$ a tower of string modes (obtained instead from three NS5-branes) \cite{Calderon-Infante:2026zmd}.  For theories with an Einstein gravity dual at low energies the exponential rate is always $\alpha=1/\sqrt{2}$, plus corrections subleading in $N$. Rewriting $\dim G_{\mathrm{free}}$ in  \eq{alpha} in terms of the central charges $a$ and $c$ (see \cite{Calderon-Infante:2024oed}), 
\begin{equation}
\label{alphaac}
    \alpha=\frac1{\sqrt{2(2a/c-1)}}
\end{equation}
one finds that $\alpha=1/\sqrt{2}$ if and only if  $a=c$ to leading order in $N$, a property necessary for the existence of an Einstein holographic dual \cite{Henningson_1998, Myers_2010}. Notice that the rewriting in terms of the central charges in \eqref{alphaac} is only valid in the overall weak-coupling limit of a SCFT. These results were generalized in \cite{Calderon-Infante:2026rkj,Calderon-Infante:2026zmd,Baume:2026svx} for certain quiver gauge theories with multiple gauge factors. In particular, it was checked in \cite{Calderon-Infante:2026rkj} that $\alpha\geq 1/\sqrt{2}$ for SU$(N)$ quiver $\mathcal{N}=2$ gauge theories (and very recently for general $\mathcal{N}=2$ quivers in \cite{Baume:2026svx}).

At first sight, the value $\alpha=1/\sqrt{2}$ appears to be in conflict with the expectation that the Sharpened Distance Conjecture gets saturated by the tower of modes of the critical ten-dimensional string,  which for a $D=5$ theory would yield $\alpha=1/\sqrt{D-2}=1/\sqrt{3}$. However, the latter expression applies only in the Einsteinian regime and therefore cannot be directly extrapolated to a small gauge-coupling limit, in which the tension of the lightest string is comparable to the spacetime curvature scale. As explained in detail in \cite{Calderon-Infante:2024oed}, the resolution of the conflict comes from accounting for the variation of the exponential decay rate of the critical string tower from the supergravity regime (large 't Hooft coupling) to the perturbative field theory regime (small 't Hooft coupling). Upon properly accounting for the quantization of the fundamental critical string in AdS, and projecting the resulting decay rate onto the conformal manifold, one indeed recovers the CFT result $\alpha=1/\sqrt{2}$ \cite{Calderon-Infante:2024oed}. Therefore, the SharpDC gets translated into the following bound
\begin{equation}
\alpha \geq \frac{1}{\sqrt{2}},
\label{SDCCFT}
\end{equation}
for infinite-distance limits in the conformal manifold. We remark that this comes from assuming that the tower of critical string modes saturates the SharpDC.\footnote{It can also be argued more generally if one incorporates the $N$ as an effective ``discrete'' infinite distance direction and follows the \emph{sliding} of the HS tower as one varies the 't Hooft coupling, as explained in \cite{Calderon-Infante:2024oed}. Assuming that this sliding occurs perpendicularly to the asymptotic direction defined by 't Hooft limit, as observed in all known cases, then the original SharpDC in the supergravity regime is only satisfied if the bound \eqref{SDCCFT} holds for limits in the conformal manifold.}
Equation \eqref{SDCCFT} is now a sharp, purely field-theoretic version of the Sharpened Distance Conjecture, which can be tested entirely in the arena of gauge theory, without making reference to gravity or AdS. This is similar to how the Charge Convexity Conjecture \cite{Aharony_2021} arose from the Weak Gravity Conjecture, but it is also a logically independent statement that may hold for generic CFTs if properly formulated \cite{Calderon-Infante:2026rkj}. 

In the original paper of the CFT Distance Conjecture \cite{Perlmutter:2020buo}, it was already noted that \eqref{SDCCFT} was satisfied in large classes of SCFTs, potentially providing a stronger bound than the strict inequality $\alpha>1/2$ coming from requiring that $c$ is necessarily bigger than the contribution of a single vector multiplet in \eqref{alpha} (see \cite{Perlmutter:2020buo}).\footnote{One can also obtain the weaker bound $\alpha>1/2$ from the conformal collider bounds \cite{Hofman_2008}, which imply $a/c< 3/2$.} 

The rest of this note is devoted to testing \eqref{SDCCFT} as a purely CFT bound.

\section{Testing the Sharpened Distance Conjecture in AdS/CFT}
\label{sec:checkSDC}
In this section, we are going to test the Sharpened Distance Conjecture in AdS \eq{SDCCFT} beyond the setting of \cite{Calderon-Infante:2024oed}. As explained in the previous section, this reference proved the bound \eq{SDCCFT} for all large $N$ theories including a single simple gauge factor.  We will therefore consider large $N$ $4d$ Lagrangian conformal field theories with a generic gauge group $G=\prod_{a=1}^M G_a$, where each $G_a$ is a simple factor (which can be SU$(N)$, SO$(N)$, or USp$(2N)$), and each matter field is charged under a representation $r=\bigotimes_a r_{a}$.

In any weak-coupling limit, we will have one or multiple gauge couplings $g_a$ going to zero at the same rate. From the discussion in Section \ref{sec:SDCADS}, as one approaches any of these points, there is an infinite tower of higher-spin states in the bulk theory becoming light exponentially fast in $5d$ Planck units. The exponential mass decay rate is computed from the CFT side via the formula \eq{alpha}. For a given gauge theory with gauge group $G$, the lowest value of $\alpha$ is attained by taking $G_\mathrm{free}=G$ in \eq{alpha}, which corresponds to the overall weak-coupling limit. Since \eq{SDCCFT} is a lower bound, we will restrict to this case without loss of generality.

We want to know whether the lower bound for $\alpha$ on the moduli space given by the Sharpened Distance Conjecture in AdS$_5$, $\alpha \geq1/\sqrt{2}$,
is satisfied in the large $N$ limit, for both supersymmetric theories (in Section \ref{checksusy}) and non-supersymmetric ones (in Section \ref{checknonsusy}), or under which conditions it can be violated.\\ 
We will also check the case of conformal field theories with a fixed $N$ and a growing number $M$ of gauge factors (see Section \ref{checklargefactors}).
\medskip

Given that we are interested in Lagrangian CFTs, conformality requires the vanishing of the exact $\beta$-functions for all gauge couplings $g_a$. We will, however, only impose a weaker condition --- the vanishing of the \emph{one-loop} $\beta$-function,
\begin{equation}
    \beta^{1\text{-loop}}(g_a)\overset{!}{=}0,
\label{eww}\end{equation}
and work in the strict large $N$ limit. This will give us a restriction on the allowed representations of charged matter, which we will exploit to compute the exponential rate explicitly. 
For a $4d$ $\mathcal{N}=2$ theory, \eqref{eww} is sufficient to ensure conformality at all loops \cite{Novikov:1983uc}. In less supersymmetric setups, this is not the case in general, but it will be enough to derive our main result.

In Table \ref{tablegroupth} we recall some group theory quantities for the simple groups SU$(N)$, SO$(N)$ and USp$(2N)$ that will be used in the next subsections.

\begin{table}[!ht]  
\centering
\renewcommand{\arraystretch}{1.3}
\begin{tabular}{ccccc}
Group & Representation $r$ & $\dim(r)$ & $T(r)$ & $C_2(r)$ \\ \hline\hline

\multirow{4}{*}{SU$(N)$}
& Fundamental 
& $N$ 
& $\tfrac12$ 
& $\dfrac{N^2-1}{2N}$ \\

& Adjoint 
& $N^2-1$ 
& $N$ 
& $N$ \\

& $2$-index Symmetric 
& $\dfrac{N(N+1)}{2}$ 
& $\dfrac{N+2}{2}$ 
& $\dfrac{(N-1)(N+2)}{N}$ \\

& $2$-index Antisymmetric 
& $\dfrac{N(N-1)}{2}$ 
& $\dfrac{N-2}{2}$ 
& $\dfrac{(N+1)(N-2)}{N}$ \\

\hline\hline

\multirow{3}{*}{SO$(N)$}
& Vector 
& $N$ 
& $1$ 
& $\dfrac{N-1}{2}$ \\

& Adjoint ($2$-index Antisymmetric) 
& $\dfrac{N(N-1)}{2}$ 
& $N-2$ 
& $N-2$ \\

& $2$-index Symmetric (traceless) 
& $\dfrac{N(N+1)}{2}-1$ 
& $N+2$ 
& $\sim N$ \\

\hline\hline

\multirow{3}{*}{USp$(2N)$}
& Fundamental 
& $2N$ 
& $\tfrac12$ 
& $\dfrac{2N+1}{4}$ \\

& Adjoint ($2$-index Symmetric) 
& $N(2N+1)$ 
& $N+1$ 
& $N+1$ \\

& $2$-index Antisymmetric (traceless) 
& $N(2N-1)-1$ 
& $N-1$ 
& $\sim N$

\end{tabular}
\caption{Low-index representations of SU$(N)$, SO$(N)$ and USp$(2N)$. The dimension, Dynkin index, and quadratic Casimir are written down for each representation. The symbol $\sim$ refers to a leading order approximation in $N$.}
\label{tablegroupth}
\end{table}

\subsection{Proof for \texorpdfstring{$4d\, \mathcal{N}\geq1$}{4d N=1} gauge theories at large \texorpdfstring{$N$}{N}}  \label{checksusy}
We consider the supersymmetric case first, by producing a sufficient set of conditions under which \eq{SDCCFT} can be derived for theories with multiple gauge factors. We will describe the matter content of $\mathcal{N}\geq1$ $4d$ SCFTs in terms of $n_r$ $\mathcal{N}=1$ chiral multiplets charged under the representation $r$ of the gauge group $G$.  We also have a number $M$ of simple gauge factors, which can be of USp, SO or SU type, and we will assume they all have the same rank $N$, so that
\begin{equation}
    \dim G = \sum_{a=1}^M\dim G_a \sim \sum_{a=1}^Md_aN^2 ,\label{dim00}
\end{equation}
at leading order in the large $N$ expansion. The coefficient $d_a$ is $1$ for SU, $\frac{1}{2}$ for SO, and $2$ for USp.

What we have described is not the most general situation, since we could also have some other gauge factors whose dimensions scale with $N$ more slowly. For instance, we could have some factors whose rank grows as $\sqrt{N}$. Our conclusions will hold as long as the total number of gauge factors with these properties gives a subleading $N$ correction to \eq{dim00}, which will always happen since the number of factors is taken to be finite and independent of $N$ in this subsection; see later in the paper for the case where this assumption is dropped. Moreover, our final result also holds in the more general case where one has gauge factors with $d_a$ different from the previous three values, for instance $G_a=$ SU$(qN)$, where $d_a=q^2$.

We start by imposing the vanishing of the supersymmetric NSVZ one-loop $\beta$-function\footnote{The NSVZ $\beta$-function at one loop, for the physical gauge coupling $g_a$, is 
\begin{equation}
    \beta_{\text{NSVZ}}^{1\text{-loop}}(g_a) = -\frac{g_a^3}{16\pi^2}\,\frac{3h_{G_a}^\vee-\sum_r n_r T_a(r)[1-\gamma_r]}{1-\frac{g_a^2}{8\pi^2}h_{G_a}^\vee}
\label{nsvz}\end{equation} 
and it is perturbatively exact for $\mathcal{N}>1$.} for each subfactor of $G$. In the large $N$ limit this corresponds to requiring\footnote{The anomalous dimensions $\gamma_r$ of the matter fields in \eqref{nsvz} have been neglected in \eqref{nsvzbcond}, since we are in the weakly-coupled limit \cite{MARTIN_1998}.}
\begin{equation} \label{nsvzbcond}
    3h_{G_a}^\vee - \sum_r n_r \, T_a(r) \overset{!}{=}0,  \quad\quad \forall a.
\end{equation}
Here $T_a(r)=T(r_{a})\prod_{a'\neq a}\dim r_{a'}$, with $T(r_{a})$ being the Dynkin index of the representation $r_{a}$ of the simple gauge factor $G_a$. We take the normalization $T=\frac{1}{2}$ for the fundamental of SU$(N)$, and recall that the dual Coxeter number $h_{G_a}^\vee = T(\text{adj})$.\footnote{Notice that for U$(1)$ factors, for which the adjoint representation is trivial, there is no way to satisfy the NSVZ vanishing condition with charged matter, which is why the group $G$ is semisimple.} 

Note that, for \eqref{nsvzbcond} to hold, both sides must scale with the same power of $N$. We have that, for the groups of our interest, $h_{G_a}^\vee\sim N$; therefore, we can only consider representations where $T_a(r)\leq3h_{G_a}^\vee$ in the large $N$ limit.\footnote{We are assuming here that the matter consists entirely of free fields when the gauge coupling vanishes exactly; dropping this assumption leads to class $\mathcal{S}$ theories, discussed later in the text.} From Table \ref{tablegroupth}, we see that this amounts to the following possibilities for each representation $r$:
\begin{itemize}
    \item Bifundamental representation (also called bif below); we will call $B_{ab}$ the number of $\mathcal{N}=1$ chiral multiplets charged under fundamentals of $G_a$ and $G_b$ (so $B_{ab}=B_{ba}$ and $B_{aa}=0$).
    \item Fundamental/Vector representation (fund); in this case we consider $F_a=F_a^{(0)}+NF_a^{(1)}$ as the number of $\mathcal{N}=1$ chiral multiplets charged under $G_a$.
    \item Index-two representation: e.g. adjoint (adj), symmetric (sym), antisymmetric (asym); we will call $J_a$, $S_a$, $A_a$ their respective number of $\mathcal{N}=1$ chiral multiplets charged under $G_a$. 
\end{itemize}
In fact, for higher indices representations, both their dimension and Dynkin index scale with higher powers of $N$. Notice that, for a complex representation $r$, the Dynkin index of the conjugate representation $r^*$ satisfies $T(r^*)=T(r)$, so there is no need to consider e.g. antifundamental separately.

Let us compute the decay rate $\alpha$ from \eqref{alpha}. Supersymmetry relates the central charge to the number $N_V$ of gauge fields and the number $N_C$ of components of matter fields (sitting in $\mathcal{N}=1$ chiral multiplets) as \cite{Anselmi_1998, Intriligator_2003}
\begin{equation}    \label{csusy}
    c = \frac{N_V}{8} + \frac{N_C}{24}= \frac{\dim G}{8} +\frac{1}{24}\sum_r n_r\dim r.
\end{equation}
Thus we can express
\begin{equation}    \label{alphasusy}
    \alpha = \frac{1}{2}\sqrt{1 + \frac{1}{3}\frac{N_C}{N_V}},
\end{equation}
and write the Sharpened Distance bound \eqref{SDCCFT} as
\begin{equation}    \label{SDCond}
    \alpha\geq\frac{1}{\sqrt{2}} \qquad\iff\qquad \frac{N_C}{N_V}\equiv\sum_r n_r\,\frac{\dim r}{\dim G}\geq 3.
\end{equation}

We want to evaluate the Sharpened Distance bound in \eqref{SDCond} explicitly, to see whether it can be violated in the large $N$ limit. By taking into account the restriction on matter fields that we obtained earlier, we will first write down how the vanishing of the one-loop $\beta$-function  \eqref{nsvzbcond} reads by considering only the allowed contributions.
It is important to observe that $\frac{h_{G_a}^\vee}{N}=1+\mathcal O\left(\frac{1}{N}\right)$ for all the three simple groups considered; then the NSVZ condition \eqref{nsvzbcond} at leading order in $N$ reads as 
\begin{equation}
\begin{aligned}
    3 + \mathcal O\left(\frac{1}{N}\right) = \frac{1}{N}\sum_r n_r \, T_a(r) &= 
    \frac{1}{N}\Bigg(\sum_{b\neq a}B_{ab}\,T_{a}(\text{bif}) + NF_a^{(1)}\,T_{a}(\text{fund}) + J_a\,T_{a}(\text{adj}) +\\
    &+S_a\,T_{a}(\text{sym}) + A_a\,T_{a}(\text{asym})\Bigg) + \mathcal O\left(\frac{1}{N}\right).
\end{aligned}
\end{equation}
Note that $T_{a}(\text{bif})=T_a(\text{fund})\cdot\dim(\text{fund}_b)$; we also define $B_a:=\sum_{b\neq a}B_{ab}\,\frac{\dim(\text{fund}_b)}{N}$. From the previous equation we then get
\begin{equation} \label{NSVZmixedcond}
    \frac{1}{N}\left[NB_a\,T_{a}(\text{fund}) + J_a\,T_{a}(\text{adj}) + S_a\,T_{a}(\text{sym}) + A_a\,T_{a}(\text{asym})\right] = 3- F_a^{(1)}\,T_{a}(\text{fund}).
\end{equation}
Now we study the LHS of the inequality in \eqref{SDCond}; for this, it is useful to recall the relation between the Dynkin index and the quadratic Casimir, $C_2$ for a simple group
\begin{equation}
    \frac{T(r_{a})}{C_2(r_{a})}=\frac{\dim r_{a}}{\dim G_a}.
\end{equation}
Using $T_a(r)=T(r_{a})\prod_{a'\neq a}\dim r_{a'}$ we find 
\begin{equation}    \label{reprtodynkin}
    \dim r=\frac{\dim G_a}{C_2(r_{a})}\,T_a(r)\sim c_a\,d_a\,N\,T_a(r) ,
\end{equation}
where the last approximate equation is valid for the representations at hand, in the large $N$ limit. We have $c_a=2$ if $r_{a}$ is fundamental, and equal to one otherwise.
Therefore, we can expand the sum in \eqref{SDCond} as
\begin{equation}
\begin{aligned}
    \frac{N_C}{N_V}\equiv\sum_r n_r\,\frac{\dim r}{\dim G} &=
    \frac{N}{\sum_ad_aN^2}\sum_{a=1}^Md_a\bigg[2N\,\frac{B_a}{2}\,T_{a}(\text{fund})+2NF_a^{(1)}\,T_{a}(\text{fund})+J_a\,T_{a}(\text{adj})+\\ &\qquad\qquad\qquad\qquad S_a\,T_{a}(\text{sym}) 
    +A_a\,T_{a}(\text{asym})\bigg]+ \mathcal O\left(\frac{1}{N}\right).
\end{aligned}
\end{equation}
We can substitute equation \eqref{NSVZmixedcond} inside the square brackets of the previous equation, to finally get
\begin{equation}    \label{resultsusy}
    \begin{aligned}
        \frac{N_C}{N_V}
        &=3 + \frac{\sum_a F_a^{(1)}\,T_{a}(\text{fund})\,d_a}{\sum_a d_a} + \mathcal O\left(\frac{1}{N}\right).
    \end{aligned}
\end{equation}
Inserting it back in equation \eqref{alphasusy}, we can compute the exponential rate, at leading order in $N$, for the theory of interest. If we consider the large $N$ $4d$ SCFTs with a simple gauge group, we recover the three universality classes classified in \cite{Calderon-Infante:2024oed} and reviewed in Section \ref{sec:SDCADS}: $\alpha = \frac{1}{\sqrt{2}}, \sqrt{\frac{7}{12}}, \sqrt{\frac{2}{3}}$.

In the more general case we focus on this paper, we notice from \eqref{resultsusy} that, at leading order in $N$, the SharpDC bound $\alpha\geq 1/{\sqrt{2}}$ is always satisfied for weakly-coupled $\mathcal{N}\geq1$ gauge SCFTs. The bound is saturated (at this order) when the number of chiral multiplets in fundamental representations does not grow with $N$, i.e. $F_a^{(1)}=0$ $\forall a$. At the next order, $1/N$, negative contributions may also appear, a phenomenon that was already noticed in \cite{Calderon-Infante:2024oed}. As a concrete example, for $G=$ SU$(N)^M$, one finds the subleading term
\begin{equation}\frac{\sum_{a=1}^M[F_a^{(0)}-(S_a-A_a)]}{2MN},\end{equation}
which can be negative. Hence, the bound is only satisfied to leading order in $N$, as expected.

We can also determine an upper bound for the exponential rate in the overall weak-coupling limit under consideration. In particular, by inspection of \eqref{NSVZmixedcond}, we notice that each term $F_a^{(1)}\,T_{a}(\text{fund})$ appearing in the sum in \eqref{resultsusy} can be at most equal to three. Therefore, equation \eqref{resultsusy} implies that $3\leq\frac{N_C}{N_V}\leq 6$, and in turn we find the following window of allowed values of $\alpha$:
\begin{equation}
    \frac{1}{\sqrt{2}}\leq \alpha \leq \frac{\sqrt{3}}{2} ,
    \label{boundalpha}
\end{equation}
at leading order in $N$, for weakly-coupled SCFTs with finitely many gauge factors.

We end this subsection by noting that, using the relations between the numbers of fields and the central charges of a $4d$ SCFT (namely equation \eqref{csusy} and $a=\frac{3N_V}{16}+\frac{N_C}{48}$), the exponential rate for supersymmetric theories in the overall weak-coupling limit can also be written solely in terms of the conformal anomaly coefficients $a$ and $c$ \cite{Calderon-Infante:2024oed}, as mentioned in \eqref{alphaac},
\begin{equation}
    \alpha = \frac{1}{\sqrt{2}}\frac{1}{\sqrt{\frac{2a}{c}-1}}.
\label{aformulac}\end{equation}
The inequality \eqref{SDCond} simply becomes $c\geq a$ at leading order in $N$. Interestingly, this behavior was noted in \cite{Parnachev_2009} on a case-by-case basis of examples of large $N$ $4d$ $\mathcal{N}=1$ interacting conformal gauge theories. Now we have proved it for all Lagrangian $4d$ SCFTs with finitely many gauge factors. Similarly, the inequality we found means that in these limits, the Sharpened Distance Conjecture predicts the Gauss--Bonnet higher-derivative coupling, whose coefficient is proportional to $(a-c)$ \cite{Blau_1999, NOJIRI_2000, Kats_2009}, has a definite sign \cite{Buchel_2010}. Perhaps interestingly, the sign we found in large $N$ theories is the one preferred by positivity of scattering amplitudes in flat space \cite{Cheung_2017}.  We remark, however, that these are two very different contexts, and that the analysis in flat space in that reference does not directly apply to AdS. It would be interesting to determine if this connection is accidental or somehow meaningful.

In terms of the ratio between the central charges of $4d$ SCFTs, the bounds on $\alpha$ in \eqref{boundalpha} become
\begin{equation}
    \frac56\leq\frac{a}{c}\leq1.
\end{equation}
The conformal collider bounds of \cite{Hofman_2008} are
\begin{equation}\frac12\leq\frac{a}{c}\leq \frac32.
\end{equation}
We observe that the bounds we have derived are more restrictive.

\subsection{Non-supersymmetric gauge theories}    \label{checknonsusy}
Although large $N$ non-supersymmetric CFTs are difficult to construct, and are in any case not expected to have exact moduli spaces, the proof  of \eqref{SDCCFT} in the previous subsection only made use of the one-loop $\beta$-function. This can be studied directly for non-supersymmetric gauge theories, and we will now repeat the analysis in the non-supersymmetric context. Consider a non-supersymmetric $4d$ conformal field theory with a gauge group $G$ with $M$, independent of $N$, mixed factors all scaling equally with $N$. Again, each factor is one among SU, SO, USp.\\
In the absence of supersymmetry, the central charge $c$ at the free point is given by \cite{Osborn_1994}
\begin{equation}
    c= \frac{1}{60}N_S + \frac{1}{40}N_F + \frac{1}{10}N_V,
\end{equation}
where $N_S$ is the number of complex scalar fields, $N_F$ is the number of Weyl fermions and $N_V$ the number of gauge fields in the theory.
Substituting the previous expression for the central charge in \eqref{alpha}, and using that $N_S=\sum_r n_{S,r}\dim r$, $N_F=\sum_r n_{F,r}\dim r$ and $N_V=\dim G= \sum_a \dim G_a$, we get this explicit form for the exponential rate:
\begin{equation}    \label{alphanonsusy}
    \alpha = \sqrt{\frac{1}{5}\left(1 + \frac{\frac{N_S}{3}+\frac{N_F}{2}}{2N_V}\right)}.
\end{equation}
The Sharpened Distance bound for the non-supersymmetric case translates into the following inequality
\begin{equation}    \label{SDCondnonsupersymmetric}
    \alpha\geq\frac{1}{\sqrt{2}} \quad \iff \quad
    \frac{N_S+\frac{3}{2}N_F}{N_V}\equiv\sum_r \left(n_{S,r}+\frac{3}{2}n_{F,r}\right)\frac{\dim r}{\dim G}\geq 9.
\end{equation}
The vanishing of the one-loop (ordinary) $\beta$-function reads for each factor as\footnote{The $\beta$-function for the gauge coupling $g_a$ at one-loop order in perturbation theory  is \begin{equation}
 \beta^{\text{1-loop}}(g_a) = - \frac{g_a^3}{16\pi^2}\left[\frac{11}{3}h_{G_a}^\vee-\sum_r \left(\frac{1}{3}n_{S,r}+\frac{2}{3}n_{F,r}\right)T_a(r)\right] .  
\end{equation}}
\begin{equation}    \label{ordbetacond}
    \beta^{\text{1-loop}}(g_a)\overset{!}{=}0 \quad\iff\quad 
    11h_{G_a}^\vee = \sum_r\left(n_{S,r}+2n_{F,r}\right)\,T_a(r),
\end{equation}
where $n_{S,r}$ and $n_{F,r}$ are the number of complex scalars and Weyl fermions in the representation $r$ of $G$, respectively. 
Again, by matching the scaling for large $N$ of both sides in \eqref{ordbetacond}, we find the same conditions as for the supersymmetric case in terms of the allowed matter; that is, each $r$ is one among bifundamental, fundamental with a number of charged fields growing with $N$, or an index-two representation. We re-emphasize that vanishing of the one-loop $\beta$-function is a necessary condition for the existence of a CFT, but in general not sufficient, so most of the theories considered here are at best dual to ``nearly AdS'' theories, which of course are not expected to be Einstein in the semiclassical (large $N$) limit.

The computation is carried out in a similar fashion to Section \ref{checksusy} if we rewrite \eqref{ordbetacond} as 
\begin{equation}    \label{ordbetacondrewrite}
    11h_{G_a}^\vee = \sum_r\left(n_{S,r}+2n_{F,r}\right)T_a(r)=: 
    \sum_r\Tilde{n}_r \, T_a(r).
\end{equation}
In this way we get an equation analogous to \eqref{NSVZmixedcond}, now with tilded quantities, which reads as
\begin{equation}\label{nonsusymixedcond}
    11 - \Tilde{F}_a^{(1)}T_a(\text{fund}) = \frac{1}{N}\left(N\Tilde{B}_a\,T_a(\text{fund}) + \Tilde{J}_a\,T_a(\text{adj}) + \Tilde{S}_a\,T_a(\text{sym}) + \Tilde{A}_a\,T_a(\text{asym})\right),
\end{equation}
at leading order in $N$. Moreover, in equation \eqref{SDCondnonsupersymmetric} we see that $n_{S,r}+\frac{3}{2}n_{F,r}=\Tilde{n}_{r}-\frac{1}{2}n_{F,r}$, by definition of $\Tilde{n}_{r}$. Using the relation between the Dynkin index and the quadratic Casimir in equation \eqref{reprtodynkin}, we can proceed analogously to the supersymmetric case (to which we refer for more details). Namely, we can expand the sum in \eqref{SDCondnonsupersymmetric} in terms of the allowed matter contributions as
\begin{equation}
\begin{aligned}
    \sum_r\left(\Tilde{n}_{r}-\frac{1}{2}n_{F,r}\right)\frac{\dim r}{\dim G} &=
    \frac{N}{\sum_ad_aN^2}\sum_{a=1}^Md_a\bigg[2N\,\frac{\Tilde{B}_a}{2}\,T_{a}(\text{fund})+2N\Tilde{F}_a^{(1)}\,T_{a}(\text{fund})+\\ &+\Tilde{J}_a\,T_{a}(\text{adj})+\Tilde{S}_a\,T_{a}(\text{sym}) 
    +\Tilde{A}_a\,T_{a}(\text{asym})\bigg]-\frac{1}{2} \frac{N_F}{N_V}+ \mathcal O\left(\frac{1}{N}\right),
\end{aligned}
\end{equation}
where 
\begin{equation}\frac{N_F}{N_V}\equiv \sum_r n_{F,r}\frac{\dim r}{\dim G}\label{ennng}\end{equation} and the coefficients $d_a$ were defined in \eqref{dim00}.
We substitute in the equation above the constraint given by \eqref{nonsusymixedcond}, and finally obtain
\begin{equation}    \label{partialresnonsusy}
    \sum_r\left(\Tilde{n}_{r}-\frac{1}{2}n_{F,r}\right)\frac{\dim r}{\dim G} = 11 + \frac{\sum_a \Tilde{F}_a^{(1)}\,T_{a}(\text{fund})\,d_a}{\sum_a d_a} -\frac{1}{2} \frac{N_F}{N_V} + \mathcal O\left(\frac{1}{N}\right) .
\end{equation}
To simplify the notation, let us call the LHS of \eqref{partialresnonsusy} as $Y$, and define $X:=\sum_r \Tilde{n}_{r}\frac{\dim r}{\dim G}$. From \eqref{partialresnonsusy}  and \eqref{ennng} we see that $X=11 + \frac{\sum_a \Tilde{F}_a^{(1)}\,T_{a}(\text{fund})\,d_a}{\sum_a d_a}$ at leading order in $N$, and equation \eqref{partialresnonsusy} is equivalently rewritten as
\begin{equation}    \label{partialresnonsusybis}
    Y=X-\frac{1}{2}\frac{N_F}{N_V}.
\end{equation}
Since we are interested in a lower bound for $Y$, we consider vanishing $\Tilde{F}_a^{(1)}$ and we need to find the maximal fraction $N_F/N_V$. By inspection of equation \eqref{ordbetacondrewrite}, we notice that $2n_{F,r}\leq \Tilde{n}_{r}$ and thus
\begin{equation}
    0\leq\ \frac{N_F}{N_V}\leq \frac{1}{2} \sum_r\Tilde{n}_{r}\,\frac{\dim r}{\dim G}\equiv \frac{X}{2} ,
\end{equation}
with the upper bound reached in the case of CFTs with no scalars. Substituting in \eqref{partialresnonsusybis}, we can conclude that $Y\geq\frac{3}{4}X$ and, since $X=11$ at leading order when $\Tilde{F}_a^{(1)}=0$, the final result is
\begin{equation}
    Y  \geq \frac{33}{4}=8.25,
\end{equation}
to be compared with \eqref{SDCondnonsupersymmetric}. In terms of the mass decay rate this means, using equation \eqref{alphanonsusy}:
\begin{equation}    \label{actualnonsusybound}
    \alpha \geq \sqrt{\frac{19}{40}} .
\end{equation}
Hence, we observe that, with the same assumptions as above (fixed number of gauge factors, and Lagrangian at weak coupling so there is no strongly-coupled matter), the bound \eq{SDCCFT} can be violated at large $N$ in non-supersymmetric field theories with vanishing one-loop $\beta$-function, which instead satisfy the less strict bound given by equation \eqref{actualnonsusybound}.\footnote{Note that \eqref{actualnonsusybound} is still bigger than $1/\sqrt{3}$.}  An example of a theory saturating \eqref{actualnonsusybound} and violating \eq{SDCCFT} is that of $11$ Weyl fermions in any index-two representation of $G_a=$ SU$(N)$ and $N_V$ gauge fields. 

Again, we remark that we only studied simple necessary conditions for conformality. While it is interesting that these conditions are no longer sufficient to enforce the Sharpened Distance Conjecture in the non-supersymmetric case, the fact that the gauge coupling will run at higher loops means that these observations must at least be taken with a grain of salt.\footnote{Perhaps with the whole salt shaker.}

\subsection{Stranger strings:  SCFTs with a large number of factors} \label{checklargefactors}
So far, we have studied \eq{SDCCFT} in supersymmetric and non-supersymmetric Lagrangian theories, with a finite number of gauge factors and free index-two matter, being the latter condition required to obtain a Lagrangian theory at weak gauge coupling. We now explore what happens when one or both of these assumptions is dropped. 

We will first see what happens when one considers matter other than free bifundamentals. Free matter charged under a factor $G_a$ can be replaced by any interacting SCFT with $G_a$ global symmetry \cite{rastelli2014superconformalindextheoriesclass, Tachikawa_2015}. As the paradigmatic example of this, we will consider $4d$ $\mathcal{N}=2$ SCFTs of Class $\mathcal{S}$ without punctures, first constructed in \cite{Gaiotto_2012}, whose gravity dual was described in \cite{gaiotto2009gravitydualsn2superconformal}. In these theories, the matter is a certain interacting superconformal field theory $T_N$, which has global symmetry SU$(N)^3$.  For any integer $g>1$, one can take $2(g-1)$ copies of the $T_N$ theory, and introduce $M=3(g-1)$ gauge factors coupled to this global symmetry, in the pattern described in \cite{gaiotto2009gravitydualsn2superconformal}, to produce a superconformal field theory. Each SU$(N)$ gauge factor is coupled to two $T_N$ blocks. We will discuss this family of theories and their holographic dual in more detail in Section \ref{reviewclassS}; for now, we check that basic conformality is satisfied and compute the mass decay rate of the tensionless string towers at infinite distance, for both the large $N$ and large $M$ cases.

First of all, we check that the NSVZ condition (equation \eqref{nsvzbcond}) is met for these theories. Each $T_N$ block contributes as $N$ hypermultiplets. Hence, for each SU$(N)$ factor one has 
\begin{equation}
    \sum_r n_r\,T(r) = \underbrace{2}_\text{\# of $T_N$}\cdot \underbrace{2N}_\text{$N$ hypers$=2N$ chirals}\cdot \underbrace{\frac{1}{2}}_\text{fundamental} + \underbrace{N}_\text{1 vector} = 3N,
\end{equation}
which equals $3h_{\text{SU}(N)}^\vee$, hence these theories are conformal.

The effective number of $\mathcal{N}=2$ vector multiplets and hypermultiplets for the Class $\mathcal{S}$ theory is obtained by summing up the contributions of all $T_N$ blocks and those of the gauge sectors and is given by \cite{gaiotto2009gravitydualsn2superconformal}
\begin{equation}
    n_v=\frac{M}{9}\left(4N^3-N-3\right),
    \qquad\qquad 
    n_h=\frac{4M}{9}\left(N^3-N\right).
\end{equation}
The central charge $c$ is
\begin{equation}
    c=\frac{n_v}{6}+\frac{n_h}{12}=\frac{M}{18}\left(2N^3-N-1\right).
\label{cclassS}\end{equation}
If we  send $k\leq M$ gauge factors to zero coupling, the decay rate in equation \eqref{alpha} is given by
\begin{equation}
    \alpha=\sqrt{\frac{M(2N^3-N-1)}{9k(N^2-1)}}\sim\sqrt{\frac{2NM}{9k}},
\label{alphaMN}\end{equation}
where in the last step we kept the leading order contribution at large $N$. In particular, in the overall weak-coupling limit, where all the $M$ gauge couplings are sent to zero, we get
\begin{equation}\label{alphalargeN}
    \alpha=\sqrt{\frac{2N^3-N-1}{9(N^2-1)}}\sim \frac{\sqrt{2}}{3}\sqrt{N},
\end{equation}
so the dependence on $M$ drops out of the equation. Although this does not violate the CFT SharpDC, we note that $\alpha$ is \emph{unbounded} as a function of $N$, a fact already noticed in \cite{Perlmutter:2020buo}. We will now use the exact expression \eqref{alphaMN} to study the large $M$ (equivalently, large $g$) limit at fixed $N$. 

The limit $M\to \infty$ at finite $N$ is interesting as it also implies that the central charge $c\to \infty$ (see \eqref{cclassS}), so that one recovers a semiclassical (i.e. weakly-coupled) gravitational EFT in the dual AdS. We already know that the SharpDC is violated for finite $N$ theories, but one could wonder if it survives as a universal property in the limit of large central charge (i.e. for weakly-coupled gravitational backgrounds) --- regardless of whether the divergence on the central charge is due to taking a large $N$ or large $M$ limit.

Remarkably, one finds that at $N=2$ (and only at this value), the CFT version of the Sharpened Distance Conjecture \eq{SDCCFT} is violated! We have in fact
\begin{equation}  \alpha_{(N=2)}=\sqrt{\frac{13}{27}}<\frac{1}{\sqrt{2}}.
\label{violation}\end{equation}
Thus, \eq{SDCCFT} is \emph{not} a general property of $4d$ $\mathcal{N}=2$ SCFTs at large central charge.  In other words, at infinite distance on the AdS moduli space (at large $M$) the lightest dual tower of higher-spin states becomes massless not as fast as predicted by the CFT Sharpened Distance Conjecture, and therefore, at a slower rate than the string modes of the critical $10d$ fundamental string in Einstein theories.

Furthermore, we can consider the generalised quiver theories constructed in \cite{Bah:2012dg}.\footnote{We thank Nikolay Bobev for pointing this out to us.} At the expense of breaking supersymmetry to $4d$ $\mathcal{N}=1$, one can introduce another
discrete parameter, $0\leq |z|\leq 1$, in the Class $\mathcal{S}$ construction. From the quiver perspective, $z$ measures the imbalance between $T_N$ blocks glued via $\mathcal{N}=2$ vector multiplets with respect to $\mathcal{N}=1$ ones, as described in \cite{Bah:2012dg}.
The conformal manifold does not get lifted by quantum corrections and gets actually enhanced by the presence of flavour symmetry Wilson line marginal couplings \cite{Bah:2012dg, Beem:2012yn}. For $z=0$ one recovers the Maldacena--Nu\~{n}ez $\mathcal{N}=1$ SCFT \cite{Maldacena:2000mw}, while the previous $\mathcal{N}=2$ Class $\mathcal{S}$ theories correspond to $|z|=1$. The central charge $c$ acquires a non-trivial $z$-dependence; in particular it is a monotonically increasing function of $|z|$. By taking the overall weak-coupling limit as above, we find that the bound \eq{SDCCFT} is violated by all the theories with $N=2$ (for any $g$, $z$), and the minimal decay rate is obtained for the case $z=0$, where $\alpha=\sqrt{29/72}$. Whenever $N>2$, instead, \eq{SDCCFT} turns out to be satisfied regardless of $z$.

Back to the $\mathcal{N}=2$ Class $\mathcal{S}$ family, the $N=2$ theory that violates the SharpDC bound in \eqref{violation} is particularly interesting: The $T_2$ theory is actually a free SU$(2)$ half-hypermultiplet in the trifundamental representation, and so the theory is actually Lagrangian --- it is the only Lagrangian Class $\mathcal{S}$ theory. In fact, it is the only possible class of  $4d$ $\mathcal{N}=2$ Lagrangian theories  involving fields in  representations higher than index two \cite{Bhardwaj_2013}. Given this, one may  be tempted to conjecture that linear or circular quivers involving only bifundamental matter might satisfy \eq{SDCCFT} in the $M\rightarrow\infty$ limit. And this is indeed true for large classes of quivers. For instance,  we have checked explicitly that the theories in \cite{Nunez_2024} all satisfy the SharpDC in the limit of a very long quiver, in accordance to the results of \cite{Calderon-Infante:2026rkj}.

However, it is not true for all such quivers. Equation \eqref{SDCCFT} can be violated even when only bifundamental fields are involved. A simple example is the orthosymplectic circular quiver of \cite{Kang_2021}. Using the explicit expressions there, together with \eqref{aformulac}, one gets for such a quiver with $M$ SO$(2N)$ gauge factors,  
\begin{equation}\alpha=\frac{1}{\sqrt{2}}\frac{1}{\sqrt{\frac{2\left(N(N-1)+\frac{5}{24}\right) M}{\left(N(N-1)+\frac{1}{6}\right) M}-1}}\approx  \frac{1}{\sqrt{2}}\,\sqrt{\frac{N(N-1)+\frac{1}{6}}{N(N-1)+\frac{1}{4}}}\leq \frac{1}{\sqrt{2}}\qquad\text{as}\,\,M\rightarrow\infty,\end{equation}
so that \eq{SDCCFT} is violated for any finite $N$. The violations become subleading as $N\rightarrow\infty$, which must be the case to comply with the general proof in Section \ref{checksusy}. This example shows that the already known subleading $N$ violations of \eq{SDCCFT} for finite $M$ survive in the large $M$ limit, where they provide $\mathcal{O}(1)$ corrections. Interestingly, the smallest possible value of $\alpha$ in this family is attained at $N=2$, where $\alpha^2= 13/27$, which is exactly the same value found for the $N=2$ Class $\mathcal{S}$ example (equation \eqref{violation}) in spite of the very different structure of the examples. Following \cite{Calderon-Infante:2024oed}, one expects that the value of $\alpha$ in the overall weak-coupling limit is tied to the microscopic nature of the dual string becoming tensionless. This would suggest that both theories, despite their apparent different structure, should share the same dual closed string background. It would be interesting to confirm this expectation.

It would also be interesting to check whether $\alpha\geq \sqrt{13/27}$ might be the actual universal lower bound for $4d$ $\mathcal{N}=2$ SCFTs at large central charge or, if it is not, find the actual minimum. At any rate, this value corresponds to a lower exponential rate than the $\alpha=1/\sqrt{2}$ obtained in the Einsteinian examples where a perturbative fundamental string becomes light in the dual AdS. 

So, what does the violation of \eqref{SDCCFT} tell us? Since we have proven that the bound holds in the standard large $N$ limit, we know that these limits of large central charge cannot correspond to ordinary large $N$ limits. In fact, it is known that quiver gauge theories where the number of factors become large provide examples of the deconstruction phenomenon \cite{Arkani_Hamed_2001, Arkani_Hamed_2003}, where the gauge theory effectively grows one extra dimension. In the bulk, this corresponds to a ``partial decompactification'' limit, where part of the internal manifold becomes large, causing the lower-dimensional Newton's constant to go to zero and the central charge of the dual CFT to diverge linearly in $M$. Taking $M$ parametrically large therefore corresponds to having part of the internal geometry much larger than the AdS space. Moreover, even if there is a string becoming tensionless as the gauge couplings vanish, there is no Hagedorn behavior as the number of factors $M$ becomes large.

We can see this somewhat explicitly in  the Class $\mathcal{S}$ example, whose holographic dual for large $N$ is well-known, and will be described in some detail in the next Section \ref{reviewclassS}. As usual, the hierarchy between the $5d$ Planck length $\ell_5$ and $5d$ AdS lengthscale $\ell_\mathrm{AdS_5}$ is given by the central charge \eqref{cclassS},
\begin{equation} \left(\frac{\ell_\mathrm{AdS_5}}{\ell_{5}}\right)^3\propto N^3 (g-1),\end{equation}
so even at $N=2$ the $5d$ theory has small gravitational coupling if $g$ is very large. As we will see in the next section, the large $g$ limit can be described as decompactification of one extra dimension, since $g$ controls the genus of a Riemann surface. However, in this case, the description is somewhat unsatisfactory. As we will see in the next section, the hierarchy between the AdS scale and the 11-dimensional Planck scale $\ell_{11}$ is controlled solely by $N$; more precisely we have ${\ell_\mathrm{AdS_5}/\ell_{11}}\sim N^{1/3}$.  For $N=2$, this means that the curvature of the AdS solution is comparable to the fundamental Planck scale of the theory --- and thus, that the very notion of spacetime may not make sense at all in this background, even when $c\rightarrow\infty$. 

One might be tempted to ascribe the violation of \eq{SDCCFT} to the very non-geometric character of the solution, but the underlying reason is more subtle. The other example we discussed, the orthosymplectic circular quivers, violate \eq{SDCCFT} in the large $M$ limit for any finite value of $N$. Although we do not have the concrete holographic dual to this solution, we expect it to fall within the large class of Gaiotto--Maldacena geometries \cite{gaiotto2009gravitydualsn2superconformal, Reid-Edwards:2010vpm}; if that is the case, the parameter $N$ will control the hierarchy between $\ell_\mathrm{AdS}$ and $\ell_{11}$, so that \eq{SDCCFT} is violated even when $N$ is large but finite and 11-dimensional gravity is weakly coupled. Therefore, what controls the violation of \eq{SDCCFT} is the fact that the large central charge happens while the 11-dimensional gravitational coupling remains finite --- whether weak or strong --- and therefore, $1/N$ effects are taken into account.   Finally, although the known violations of \eq{SDCCFT} only happen in decompactification limits as above, not all such limits violate \eq{SDCCFT}. One example which respects the conjecture is the linear SU$(N)$ quivers of \cite{Nunez_2024}, as we explained above (see \cite{Calderon-Infante:2026rkj}).

In conclusion, we have shown that there are $4d$ SCFTs that violate the SharpDC bound \eqref{SDCCFT} in the limit of large central charge $c\to \infty$, although this violation disappears if in addition $N\to \infty$. Since whether \eq{SDCCFT} is violated or not amounts to whether $a> c$ or not, it is perhaps not surprising that we can find even large $c$ families of CFTs that have $a>c$ and hence violate the bound. Equivalently, we  can rephrase our result from section  \ref{checksusy} as the statement that $4d$ $\mathcal{N}\geq 1$ Lagrangian gauge theories in the large $N$ limit with vanishing one-loop $\beta$-function satisfy $a\leq c$.  It would be interesting to attempt a generalization of our proof of Section \ref{checksusy} for non-Lagrangian SCFTs like Class $\mathcal{S}$, which admit weak-coupling limits of the gauge sector but with strongly-coupled matter. This would allow us to elucidate whether \eqref{SDCCFT} is always true at large $N$ or not. 

\section{Testing the Refined Distance Conjecture in AdS/CFT}
\label{sec:checkRDC}
In the previous section, we have emphasized that the SharpDC need not be satisfied outside the strict large $N$ limit, and showed a violation involving the simplest Class $\mathcal{S}$ theory. In this section we will see how the same class of theories, now at large values of $N$,  might pose a challenge to another Swampland Conjecture of interest: The Refined Distance Conjecture of \cite{Baume_2016, Klaewer_2017, Rudelius:2023mjy}, which limits the excursion in field space that a field can have before the tower of states predicted by the Distance Conjecture sets in.  

We will first review briefly the theories of Class $\mathcal{S}$ that we will study, and then describe the Refined Distance Conjecture and how to test it in the present setup.

\subsection{Class \texorpdfstring{$\mathcal{S}$}{S} theories and their infinite distance limits}
\label{reviewclassS}
We will now review the Class $\mathcal{S}$ theories first constructed by Gaiotto in \cite{Gaiotto_2012} and whose gravity dual was described in \cite{gaiotto2009gravitydualsn2superconformal}, to which we refer the reader for more details. 

Class $\mathcal{S}$ theories are $4d$ $\mathcal{N}=2$ SCFTs, obtained by compactifying $6d$ $\mathcal{N}=(2,0)$ theories on a Riemann surface, $\Sigma_g$, of genus $g>1$ and no punctures. In this note, we will focus on the $A_{N-1}$ family in $6d$, which is described by a quiver gauge theory with group $G=$ SU$(N)^M$ in $4d$. The difference from ordinary quiver gauge theories lies in the matter sector coupled to the gauge fields. Except for the case $N=2$, in Class $\mathcal{S}$ theories  the matter sector is itself a strongly-coupled, interacting conformal field theory. This makes Class $\mathcal{S}$ theories non-Lagrangian.

The interacting SCFT providing the matter sector of these theories is the so-called $T_N$ theory \cite{Tachikawa_2015}. This is an isolated SCFT obtained by compactifying the $6d$ $\mathcal{N}=(2,0)$ on a sphere with three punctures.\footnote{The dimension of the moduli space of a Riemann surface $\Sigma_{g,n}$ of genus $g$ and $n$ punctures is $\dim_{\mathbb{C}}\mathcal{M}_{g,n}=3g-3+n$. Hence, the moduli space of a three-punctured sphere is just a point, and $T_N$ theories do not admit marginal deformations.} As mentioned above, only for the case of $T_2$ there is a Lagrangian description, corresponding to SU$(2)$ gauge fields with half-hypermultiplets in trifundamental representations \cite{Gaiotto_2012}. 

\begin{figure}[!ht]
    \centering
    \includegraphics[width=0.8\textwidth]{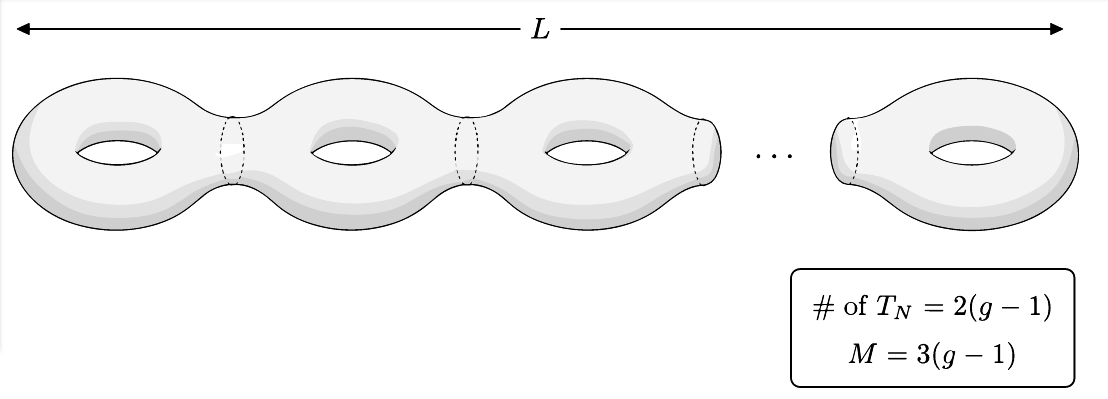}
    \caption{A genus-$g$ Riemann surface constructed by merging $g$ tori. The length of the chain is $L$, and corresponds to the extra dimension emerging in the large $g$ limit. The number of $T_N$ SCFTs blocks and the number of gauge factors $M$ in terms of the genus $g$ are also displayed.}
    \label{fig:riemann}
\end{figure}

Each $T_N$ factor has a SU$(N)^3$ global symmetry; the Class $\mathcal{S}$ theories are then obtained by introducing weakly-coupled gauge fields that gauge these global symmetries, in a pattern specified by a diagram akin to a pants decomposition of a Riemann surface of genus $g$, where one has $g$ tori connected by $g-1$ hyperbolic cylinders (see Figure \ref{fig:riemann}). The number of $T_N$ blocks is related to the genus as $2(g-1)$, and the total number of gauge factors is $M=3(g-1)$. 

In the previous section, we studied the Lagrangian example of this family, where $N=2$. We will now focus on the large $N$ limit, where there is also a dual supergravity description \cite{gaiotto2009gravitydualsn2superconformal}. Class $\mathcal{S}$ theories without punctures are dual to AdS$_5$ compactifications of M-theory on a space that topologically is a fibration $S^4\longrightarrow\Sigma_g$, where the $S^4$ is threaded by $N$ units of $G_4$ flux, and $\Sigma_g$ is equipped with its metric of constant, negative scalar curvature. The $11d$ metric reads as \cite{gaiotto2009gravitydualsn2superconformal}
\begin{equation} 
\begin{aligned}
ds_{11}^2
&= (\pi N \ell_{11}^3)^{2/3}\,W^{1/3}\Biggl\{
2\,ds^2_{\mathrm{AdS}_5} +\\
+&\left[
\frac{dr^2+r^2 d\beta^2}{(1-r^2)^2}
+ d\theta^2
+ \frac{1}{W}\cos^2\theta\left(d\psi^2+\sin^2\psi\,d\phi^2\right)
+ \frac{2}{W}\sin^2\theta\left(d\chi+\frac{2r^2\,d\beta}{1-r^2}\right)^2
\right]
\Biggr\},\\
\label{metric}
\end{aligned}
\end{equation}
with $W \equiv 1+\cos^2\theta$. Since the latter depends explicitly on the angle, in the following formulas we will fix $\theta=\pi/2$ and so $W=1$, for simplicity.

The Riemann surface, with coordinates $r$ and $\beta$, is itself specified as a quotient of the hyperbolic space $\mathbb{H}$ by a discrete subgroup $\Gamma\subset\mathrm{PSL}(2,\mathbb{R)}$. From this quotient it inherits a metric of constant negative curvature, which is the one appearing above. Using the Gauss--Bonnet theorem, we can determine the area of the embedded copy of $\Sigma_g$ at e.g. $\theta=\pi/2$, as
\begin{equation}
    \text{Area}(\Sigma_g) = 4\pi(g-1)(\pi N \ell_{11}^3)^{2/3}\,
\label{area0}\end{equation}
in $11d$ Planck units. 

The central charge of Class $\mathcal{S}$ theories was given above in \eqref{cclassS}. In the dual gravitational theory, this fixes the ratio\footnote{Equation \eqref{cclassS} was in terms of $M$, while here we write the same equation in terms of the genus $g$.}
\begin{equation}
    c=\frac{\pi}{8}\left(\frac{\ell_\mathrm{AdS_5}}{\ell_5}\right)^3=\frac{(g-1)}{6}\left(2N^3-N-1\right).\label{ccharge}
\end{equation}
As in the previous section, we observe that there are two independent ways to obtain a semiclassical (i.e. weakly-coupled) gravitational EFT in the dual AdS: either by taking a large $N$ limit or large $M$ (i.e. large genus $g$). In the latter case, \eq{area0} tells us that by increasing $g$, the area of $\Sigma_g$ increases. Picturing the Riemann surface as a series of $g$ tori glued to each other as in Figure \ref{fig:riemann}, we see that this area increase is driven merely by the addition of more and more tori, so this is a decompactification limit where one extra dimension emerges.

From the prefactor in the metric \eqref{metric}, which is in 11-dimensional Planck units, we read that
\begin{equation}
    \left(\frac{\ell_\mathrm{AdS_5}}{\ell_{11}}\right)^3\sim N.
\end{equation}
That is, the hierarchy between $\ell_\mathrm{AdS_5}$ and $\ell_{11}$ is controlled exclusively by $N$. Only at large $N$ is the solution reliably described by 11-dimensional supergravity;  the genus $g$ can be increased independently of this. Notice also that, as we vary $N$, we are moving within a family of theories with increasing spatial volume but with the same background topology; instead, when  $M$ changes, the genus of the Riemann surface increases, and so the topology of the internal manifold changes. 

\subsubsection*{Limits of large central charge}

Having described the basic features of the solution, we now discuss the different limits providing a large central charge. From the perspective of \cite{Lust:2019zwm}, they correspond to infinite distance limits in the space of metric configurations. Generalizing the Distance Conjecture to this metric space --- often referred to as the AdS Distance Conjecture --- the authors in \cite{Lust:2019zwm} proposed that such limits should also exhibit an infinite tower of states becoming light.

For $N\rightarrow\infty$, there is indeed a light tower of KK modes associated to the decompactification of the $S^4$, whose mass scale is
\begin{equation}
M_\mathrm{KK}^{S^4}\,\ell_5=\ell_\mathrm{AdS_5}^{-1}\sim\frac{1}{N(g-1)^{1/3}}.
\end{equation}
This coincides with the inverse of the sphere curvature in $5d$ Planck units,\footnote{The relation between the $5d$ Planck scale and the $11d$ one is
\begin{equation}
    \ell_5^{-3}=\frac{\text{Vol}_\Sigma\,\text{Vol}_{S^4}}{\ell_{11}^9}=\ell_\mathrm{AdS_5}^6\frac{4\pi(g-1)\,\frac{8}{3}\pi^2}{\ell_{11}^9}=\frac{32\pi^5}{3}(g-1)N^2\ell_{11}^{-3},
\end{equation}
where we used $\text{Vol}_\Sigma=4\pi(g-1)$ and $\text{Vol}_{S^4}=\frac{8\pi^2}{3}$ in the respective powers of AdS units.} showing that the geometry is not scale-separated (and agreeing with the general considerations in \cite{Lust:2019zwm}). 

We can also ask what happens when $g\rightarrow\infty$ at fixed $N$. Here we encounter the first unusual property of infinite distance limits in the Class $\mathcal{S}$ solution. As $g\rightarrow\infty$, we decompactify a single extra dimension, and therefore, we might expect to find a tower of KK modes whose size is controlled by $1/L$, where $L$ is the length of the extra dimension (the longer side of the glued chain of tori in Figure \ref{fig:riemann}). However, this expectation is wrong. KK modes of fields on $\Sigma_g$ obey Laplace's equation in this space; due to the negative scalar curvature, the spectrum of the Laplacian is in general controlled by the curvature scale of the induced metric  (as explained in the discussion around \eq{area0}), and not by $L$ (see e.g. \cite{Selberg1965Estimation,2021arXiv210205581W}). As a result, there is no tower of KK modes associated to the decompactification of an extra dimension in the $g\rightarrow\infty$ limit! 

Of course, the AdS Distance Conjecture still demands the presence of some tower becoming light, in five-dimensional Planck units, even in this limit. And indeed, KK modes of the $S^4$ do become light, since they have masses of order the $5d$ AdS scale and in the $g\rightarrow\infty$ limit the central charge of the field theory (and hence the hierarchy between $\ell_\mathrm{AdS}$ and $\ell_5$ in the bulk, as determined by \eq{ccharge}) diverges. AdS solutions which are not scale separated automatically comply with the AdS Distance Conjecture of \cite{Lust:2019zwm}. However, this is an interesting example in which such limit of large central charge does not map to an infinite distance limit in field space of the bulk dual, as the topology of the background also changes. Hence, according to the original Distance Conjecture defined in moduli spaces, there is no need to have any tower. Still, we find the KK tower, in agreement with \cite{Lust:2019zwm}, but the exponential mass decay rate is not what one would have expected from the microscopic interpretation of the limit: the KK tower is associated only to the $S^4$, even if a single extra dimension (the length of the chain) decompactifies at a faster rate in such limit.

\subsubsection*{Limits on the conformal manifold}

On top of the large $N$ and large $g$ limits we have just described, these theories also have a rich bulk moduli space, providing a conformal manifold with many infinite distance limits on which we will focus from now on.

From the gauge theory perspective, the $M=3(g-1)$ SU$(N)$ factors lead to $3(g-1)$ complexified gauge couplings, which are exactly marginal deformations in $4d$ $\mathcal{N}=2$ gauge theory and which parametrize the conformal manifold of the theory. In the bulk, these become complex structure moduli of $\Sigma_g$ \cite{gaiotto2009gravitydualsn2superconformal}. The moduli space of Riemannian metrics of constant negative curvature on $\Sigma_g$, called $\mathcal{M}_g$, is a classical object in geometry \cite{hubbardteichmuller}, and its dimension is precisely $\dim_{\mathbb{C}}\mathcal{M}_g=3(g-1)$, matching the field theory counting.

Next we need to discuss the metric on $\mathcal{M}_g$. The CFT Distance Conjecture of \cite{Perlmutter:2020buo, Baume:2020dqd} is phrased in terms of the Zamolodchikov metric, which is the natural metric on the conformal manifold of the CFT. As discussed in detail in \cite{Perlmutter:2020buo}, for an Einstein theory of gravity the Zamolodchikov metric is proportional to the metric defined by the kinetic terms of the moduli fields, which is the one in which the Distance Conjecture is usually formulated. Since we are interested in the large $N$ limit, we can use supergravity reduced on the Gaiotto--Maldacena background to compute the kinetic term of the moduli fields. Experience from Calabi-Yau compactifications suggests that the metric should then be proportional to the so-called Weil--Petersson metric on the Teichm{\"u}ller space  \cite{BuserPeter1992Gaso, wolpert2007weilpeterssonmetricgeometry}, and indeed, an explicit calculation \cite{Tachikawa_2017} shows that this is the case. More concretely, the resulting AdS moduli space metric in five-dimensional Planck units is simply\footnote{From \cite{Perlmutter:2020buo}, equation 4.5, we have the following relation between the AdS$_5$ moduli space metric and the Zamolodchikov metric on the dual $\mathcal{M}_\mathrm{CFT_4}$
\begin{equation}
\label{Zm}
    g_{I\bar J}^{\text{(mod)}} = \frac{1}{192\pi^2}\left(\frac{\ell_5}{\ell_\mathrm{AdS_5}}\right)^{3}g_{I\bar J}^{\text{(Zam)}}.
\end{equation}
Eq. 3.17 in \cite{Tachikawa_2017} gives the proportionality between the Zamolodchikov metric and the Weil--Petersson one:
\begin{equation}    \label{zamWP}
    g_{I\bar J}^{\text{(Zam)}}=24\pi^2\frac{\ell_\mathrm{AdS_5}^3}{8\pi G_N^{(7)}}\,g_{I\bar J}^{\text{(WP)}}=\frac{96\pi}{g-1}\left(\frac{\ell_\mathrm{AdS_5}}{\ell_5}\right)^{3}g_{I\bar J}^{\text{(WP)}}.
\end{equation}}
\begin{equation}    \label{modWP}
    g_{I\bar J}^{\text{(mod)}} = \frac{g_{I\bar J}^{\text{(WP)}}}{2\pi(g-1)},
\end{equation}
where $g_{I\bar J}^{\text{(WP)}}$ is the standard Weil--Petersson metric for $\Sigma_g$ with unit scalar curvature.

The relation \eqref{modWP}  is valid in the supergravity approximation. It is good at large $N$ and large 't Hooft coupling. In general, the metric of this $4d$ $\mathcal{N}=2$ conformal manifold is special K{\"a}hler \cite{Tachikawa_2017}, but is otherwise unprotected, so it can receive corrections.  These corrections are of two types. At finite $N$, the metric receives $1/N$ corrections. We will not address these in this work, and focus on large $N$ only. But even then, the metric can receive corrections. The WP metric on $\mathcal{M}_g$ is not everywhere smooth. In fact, it is geodesically incomplete, and it exhibits finite distance singularities called nodal degenerations (see \cite{wolpert2005geometry} and references therein). Geometrically, these degenerations correspond to pinching, where a locally length-minimizing curve  (called \emph{systole}) of the Riemann surface $\Sigma_g$ goes to zero size. An M2-brane wrapping this one-cycle becomes light, and is expected to give a tensionless string. Since we would identify a light string with a weak-coupling limit on the gauge theory, and these are always at infinite distance \cite{Perlmutter:2020buo, Baume:2020dqd}, we expect that, in the vicinity of nodal degenerations, the WP metric should receive significant corrections which generate an infinite distance limit. 

We will now estimate the region of moduli space where the rescaled WP metric \eqref{modWP} is a good approximation to the exact result. To do this, it is convenient to introduce Fenchel--Nielsen coordinates on $\mathcal{M}_g$ \cite{wolpert2005geometry, wolpert2007weilpeterssonmetricgeometry}, which cover the region of the Teichm{\"u}ller space close to the nodal degeneration. Close to one such degeneration, one can find a pants decomposition of  $\Sigma_g$, where the surface degenerates into a collection of long hyperbolic cylinders glued together. The Fenchel--Nielsen (FN) coordinate system $(t_I, l_I)$, with $I=1\dots 3(g-1)$ (see Section 7 in \cite{wolpert2007weilpeterssonmetricgeometry}) parametrizes the moduli space in terms of the $l_I$, the systole of each hyperbolic cylinder, while the $t_I$ account for the relative gluing angles between adjacent cylinders. In the FN coordinates, the WP K{\"a}hler form is $\omega_\text{WP}=\frac{1}{2}\sum_Idl_I\wedge dt_I$, and the WP metric restricted to the $t_I=0$ slice (which we do for ease of presentation) becomes
\begin{equation}    \label{WPmetricFNcoord}
  ds^2_{\text{WP}}\simeq 2\pi\sum_I(dl_I^{1/2})^2=\frac{\pi}{2}\sum_I\frac{dl_I^2}{l_I}.
\end{equation}
Nodal degenerations correspond to the limit where some $l_I$ becomes small \cite{PMIHES_1969__36__75_0}; we can see explicitly that they are at finite WP distance, as anticipated before.

As stated above, when $l_I\rightarrow0$ the geometry develops a long hyperbolic cylinder (see Collar Theorem \cite{BuserPeter1992Gaso}). In fact, for small $l_I$, there is a region of the Riemann surface where the metric is given by
\begin{equation}
    ds^2_\Sigma\simeq W^{1/3}(\pi N\ell_{11}^3)^{2/3}\left[d\rho^2+l_I^2\cosh^2(\rho)d\varphi^2\right],
\end{equation}
where $W\equiv(1+\cos^2\theta)$, $\varphi\in[0,1]$ and the coordinate $\rho$ takes values in between $[-a,a]$, where 
\begin{equation}
    a(l_I)=\text{arcsinh}\left(\frac{1}{\sinh(\frac{1}{2}l_I)}\right)
\end{equation}
measures the ``cylinder length''. 
In this metric, the curve at $\rho=0$ is a closed geodesic that locally minimizes the distance, i.e. a local systole
of the manifold. The physical length of the systole is given by $l_\text{sys}=l_I\,(\pi N)^{1/3}\,\ell_{11}$, in $11d$ Planck units.
We see that in the limit of vanishing $l_\text{sys}$, $a$ diverges, so the geometry develops a long, thin tube. An M2-brane wrapped around the systole (tension $T_\text{M2}=\frac{1}{(2\pi)^2\ell_{11}^3}$) will become a type IIA string whenever $l_\text{sys}\lesssim \ell_{11}$, whose mass scale is given by
\begin{equation}
M_{\text{str}}=\sqrt{T_\text{M2}\,l_\text{sys}}=\frac{1}{2\pi}\frac{l_\text{sys}^{1/2}}{\ell_{11}^{3/2}}.
\end{equation}
In this regime, we already expect corrections to the M-theory description, but as long as the string scale remains above $1/\ell_{\text{AdS}_5}$, the theory will still be Einstein --- we will have a IIA compactification, rather than an M-theory background. 

In AdS units this mass is
\begin{equation}
M_{\text{str}}\,\ell_{\text{AdS}_5}\sim \sqrt{l_IN}\sim \sqrt{\frac{l_\text{sys}}{\ell_{11}}}\,N^{1/3}.
\end{equation}
Hence, the string length becomes of order the AdS scale when 
\begin{equation}
l_\text{sys}\sim \frac{\ell_{11}}{N^{2/3}}.
\label{rrr5}\end{equation}
Using the relationship $g_s\sim (l_{\text{sys}}/\ell_{11})^{3/2}$, we see that this happens at the IIA dilaton value
\begin{equation} g_s\sim \frac{1}{N}.\label{rrr10}\end{equation}
Using the usual relationship $g_\mathrm{YM}^2=g_s$, we see this threshold corresponds precisely to 't Hooft coupling $\lambda\equiv g_\mathrm{YM}^2N\sim 1$, as expected.

Using \eqref{WPmetricFNcoord}, which gives us the WP metric in terms of systole lengths, we can determine the distance from the nodal curve when $l_\text{sys}\sim \ell_{11}$, that is, when the physical length becomes Planckian and we enter the type IIA regime, generating important corrections to the metric. This happens at a WP distance
\begin{equation}    \label{epsradius}
\varepsilon\sim N^{-1/6}
\end{equation}
from this singular locus, measured according to the WP metric. Although the M-theory picture may receive corrections, the tension of the string is still far above the AdS scale, and there is a ten-dimensional Einsteinian description of the compactification. However, at an even smaller radius
\begin{equation}\label{tildeepsradius}
\tilde{\varepsilon}\sim N^{-1/2},
\end{equation}
given by \eqref{rrr5},
the string tension reaches AdS scale. At this point there can no longer be any Einstein description.

\begin{figure}[!ht]
    \centering
    \includegraphics[width=0.5\textwidth]{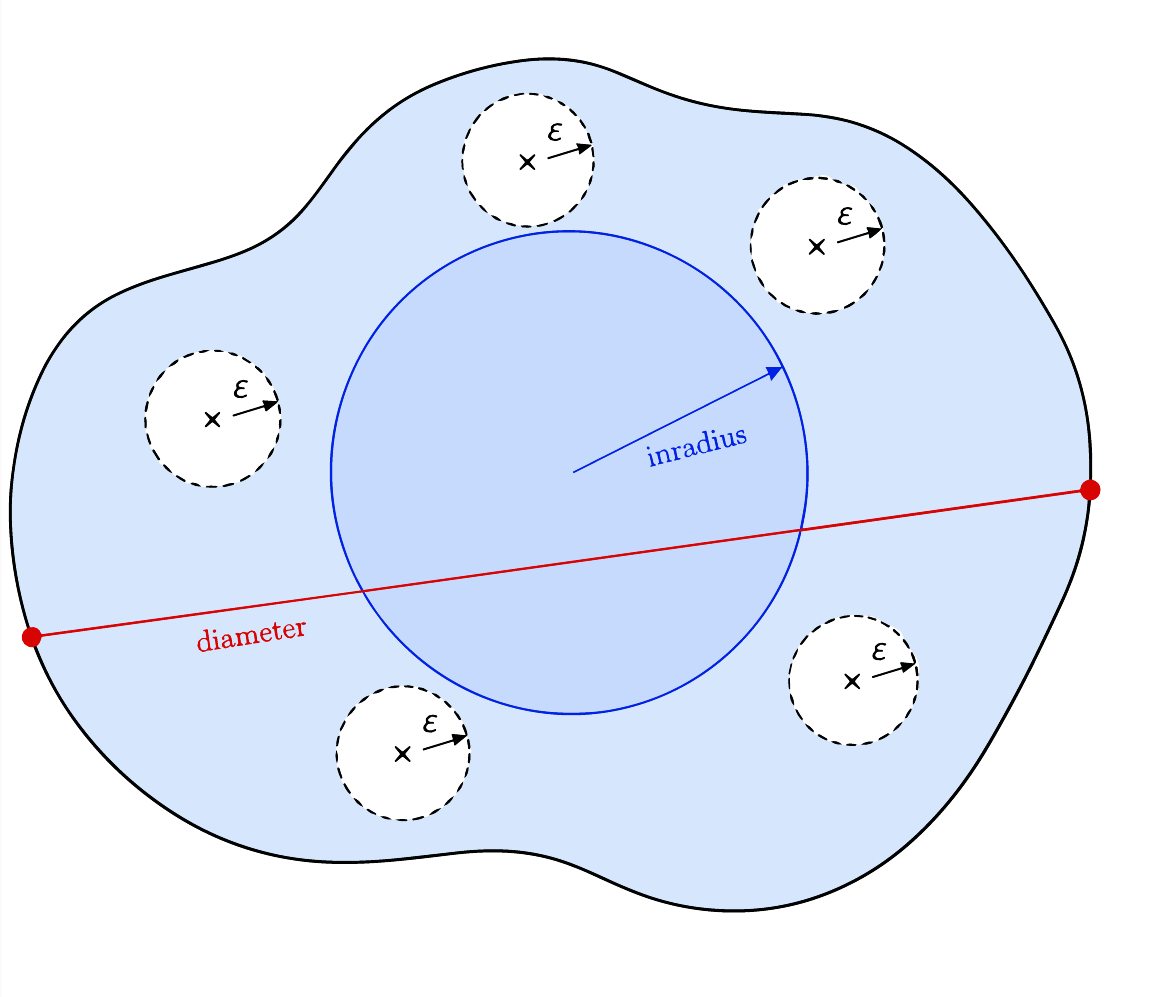}
    \caption{The moduli space of genus-$g$ Riemann surfaces $\mathcal{M}_{g}$. Little balls of radius $\varepsilon$ centered at the singularities are excised. The diameter (red) and the inradius ball (darker blue) are also shown.}
    \label{fig:modulispace}
\end{figure}

The picture of moduli space that emerges from this discussion is  shown in Figure \ref{fig:modulispace}. The moduli space of the theory, at large $N$, is the usual Teichm{\"u}ller space for constant curvature metrics on a Riemann surface of genus $g$. The space is equipped with metric \eqref{WPmetricFNcoord}, except in some small neighbourhoods of the nodal curves. Within these balls, of radius $\varepsilon$ or $\tilde{\varepsilon}$, a systole becomes sub-Planckian in the compactification manifold and the metric receives strong quantum corrections, which push the degeneration to infinite distance as we described above. 

\subsection{Testing the Refined Distance Conjecture}
\label{testAdSRDC}
The moduli space of Class $\mathcal{S}$ theories we have just described provides a unique opportunity to test the AdS version of the Refined Distance Conjecture (RDC) \cite{Hebecker_2017, Blumenhagen_2018, Rudelius:2023mjy}. This conjecture proposes that a tower of states must become light any time one traverses a specified field range in moduli space $\Delta\phi=\beta\, M_{D}$, where $\beta\sim \mathcal{O}(1)$ and $M_{D}$ is the Planck mass. It differs from the standard Distance Conjecture in that the latter only requires the tower to become light near infinite distance limits; it says the tower will become light eventually, but it does not specify exactly how far can one travel in moduli space before this happens.

For phenomenological applications, then, the Distance Conjecture is not very helpful,\footnote{Unless the inflationary model precisely takes place in the asymptotic (infinite distance) regimes of moduli space rather than in its interior. This is actually the case for most stringy-inspired models, since these asymptotic regimes are the regions where we have computational control, but it is not a priori required by nature.} since the onset of the exponential tower could take very long to appear, e.g. longer than inflation lasts. In this context, one is much more interested in a statement like the RDC, which would put a sharp upper bound to the field range that can be traversed.

Although usually described in flat space, the RDC admits a natural formulation in AdS/CFT, where it would limit the distance in the conformal manifold that can be traversed before the Einstein description of the dual AdS theory breaks down. As explained in \cite{Perlmutter:2020buo}, the field range $\beta$ in the bulk corresponds, in the dual field theory language, to a certain distance in the Zamolodchikov metric controlled by the central charge $c$. The  relation between Zamolodchikov and bulk moduli space metrics for four-dimensional SCFTs is given by \eqref{Zm} (see \cite{Perlmutter:2020buo}),
\begin{equation}
\text{Zamolodchikov distance}\sim\sqrt{c}\,\cdot\,\text{Moduli space distance},\end{equation}
up to universal numerical prefactors that will not be of relevance here.\footnote{The central charge appears to change from AdS to bulk Planck units.} Therefore, the AdS version of the RDC would naturally translate to the statement that the Zamolodchikov diameter of the region of moduli space described by Einstein gravity in the bulk should not be bigger than $\sqrt{c}$,
\begin{equation}\text{diam}^{(\text{Zam})}(\mathcal{M}_\text{CFT})\lesssim \sqrt{c}\,\beta,\label{e334}\end{equation}
where again $\beta\sim \mathcal{O}(1)$. 

This claim can be tested in the Class $\mathcal{S}$ moduli space we have just studied since, as depicted in Figure \ref{fig:modulispace}, the Einstein region of moduli space is the one outside the $\varepsilon$-balls, where the moduli space metric is proportional to the WP metric. After using \eqref{zamWP} to recast \eqref{e334} in terms of the WP metric, the factors of $N$ drop out and we get, in the large $N$ limit,\footnote{In \cite{Bah:2012dg} a large set of AdS$_5$ M-theory backgrounds were constructed, in which a discrete parameter $z$ controls the topology of the $S^4$ fibration over $\Sigma_g$ in the $11d$ supergravity description. The Gaiotto--Maldacena background described in the text corresponds to $|z|=1$. The central charge in \eqref{ccharge} gets modified, acquiring a non-trivial $z$-dependence.
However, since $c$ is absent in the final expression \eqref{rdc444}, our result is actually independent of $z$, assuming that the moduli space metric in the Einsteinian regime is still the WP one.} 
\begin{equation}\text{diam}^{\text{(WP)}}(\mathcal{M}_g) \lesssim  \sqrt{g-1}.\label{rdc444} \end{equation}
In other words, the AdS version of the Refined Distance Conjecture applied to Class $\mathcal{S}$ theories in the large $N$ limit becomes a purely mathematical conjecture: That  the  WP diameter of the moduli space $\mathcal{M}_g$ of Riemann surface of genus $g$, cannot scale faster than $\sqrt{g}$ at large $g$. \\

Mathematicians have studied the diameter of this moduli space, but the asymptotic dependence of  $\text{diam}^{\text{(WP)}}(\mathcal{M}_g)$ as a function of $g$ is not known; it is not even known whether this quantity has a well-defined limit. The most recent result we could find is that in the $g\rightarrow\infty$ limit, there are both lower and upper bounds \cite{Cavendish_2012}
\begin{equation}
\sqrt{g}\lesssim\text{diam}^{\text{(WP)}}(\mathcal{M}_g)\lesssim\sqrt{g}\log(g).
\label{bsat}\end{equation}
The AdS RDC bound \eqref{rdc444} implies then that the lower bound \eqref{bsat} must be saturated (with a not too large numerical coefficient), and that any growth of the diameter faster than $\sqrt{g}$ is impossible. Thus, in this case, Swampland principles are making a prediction about an open mathematical problem! 

It would be extremely interesting to verify or disprove this, since a resolution either way would have far-reaching implications:\begin{itemize}
\item If \eqref{rdc444} is vindicated, it would provide an impressive test of Swampland principles and their value as guides even in mathematics, not unlike the relationship between the Weak Gravity Conjecture in M-theory and the integral Hodge conjecture \cite{Heidenreich_2017}. Similarly, any mathematical techniques used in the putative proof of \eqref{rdc444} could possibly be repurposed for establishing the RDC in this and other contexts.
\item If, on the contrary, \eqref{rdc444} is shown to be false, then we would learn that we can have effective field theories which are valid for arbitrarily long field ranges, even if this would be in an AdS, non scale-separated context. Extrapolating to cosmological setups, this would mean that perhaps the RDC is not so constraining during inflation; and that we should be looking at corners of the string Landscape which share some features with Class $\mathcal{S}$ to evade it. The connection to scale separation should also be explored. In any case, if \eqref{rdc444} is false, it means that any attempt at proving the CFT Distance Conjecture of \cite{Perlmutter:2020buo, Baume:2020dqd}, by showing that states must become light after a long excursion in moduli space, must necessarily involve infinitely long field ranges (or at least, field ranges that scale with some power of $c$ in Planck units).
    
\end{itemize}

One could worry that the conclusion above is perhaps too strong; it could be that the curve realizing the diameter of $\mathcal{M}_g$ passes very close to one of the weak-coupling $\varepsilon$-balls, in which case, the EFT breaks down. However, as shown in \eqref{epsradius} and \eqref{tildeepsradius}, the radius $\varepsilon$ of the balls can be reduced arbitrarily simply by rescaling $N$. As a result, these corrections do not help in avoiding the conclusion.  Similarly, the nodal curves that correspond to singular points in the WP metric are sparse enough that the metric completion of $\mathcal{M}_g$ with the WP metric, $\overline{\mathcal{M}_g}$, has  $\mathcal{M}_g$ as a dense subset \cite{wolpert2005geometry}. This means that, even if the diameter is realized by some geodesic going through nodal curves, this can be avoided by perturbing the endpoints slightly, without affecting the diameter significantly. Therefore, finding a sequence of  values of $g$ such that the diameter grows faster than $\sqrt{g}$ would be enough to disprove the AdS version of the RDC.

Finally, it has been shown that in the $\varepsilon$-thick region of $\mathcal{M}_{g}$ (the region away from any $\varepsilon$-ball) the gap in the scalar Laplacian satisfies, at large $g$, the inequalities \cite{Wu_2022}
\begin{equation}
    \frac{\varepsilon^2}{16\pi^2g^2}\leq\min\lambda_1\leq\frac{\varepsilon}{g^2}.
\end{equation}
Since the lower bound is away from zero, this shows rigorously that, as we move in the moduli space at fixed genus, there are no unexpected towers of KK modes becoming arbitrarily light at any point along the diameter.

For completeness, in Figure \ref{fig:modulispace} we have also included the ``inradius'' of $\mathcal{M}_g$. This is the size of the largest ball that can be constructed in $\mathcal{M}_g$ without touching any degeneration; any curve within this region is, in some sense, far away from every infinite distance limit. Interestingly, the inradius grows as $\sqrt{\log g}$ \cite{AIF_2019__69_3_1309_0}, which is slow enough to avoid conflict with the RDC. However, we emphasize that restricting to a ball of inradius size is too strong a restriction: By making $\varepsilon$ small enough, the full diameter of moduli space can be considered, and hence, the tension with the RDC is real.\\ 

\textbf{Acknowledgements}: We thank Nikolay Bobev, Jos\'e Calder\'on-Infante, \'Angel M. Uranga, and Cumrun Vafa for discussions. The authors thank
the Spanish Research Agency (AEI) for support through the grants IFT Centro de Excelencia
Severo Ochoa CEX2020-001007-S, PID2021-123017NB-I00, PID2024-156043NB-I00, and CEX2025-001574-S, funded
by MCIN/AEI/10.13039/501100011033 and by ERDF A way of making Europe. The research presented in this publication falls under the research line ``Strings and Quantum Gravity'' of CEX2025-001574-S.
The work of GF is supported by the fellowship LCF/BQ/DI24/12070008 from the ”la Caixa” Foundation (ID
100010434). Throughout
the completion of this project, MM has been supported by the grants RYC2022-037545-I and EUR2024-153547 from the AEI. The work of I.V. is supported by the ERC Starting Grant QGuide101042568 - StG 2021, and the Project ATR2023-145703 funded by MCIN/AEI/10.13039/501100011033.

\bibliography{BLrefs}
\bibliographystyle{JHEP}
\end{document}